\documentclass[prd,twocolumn,floatfix,preprintnumbers,superscriptaddress]{revtex4-1}
\usepackage{amsmath}
\usepackage{graphicx}
\usepackage{color}
\usepackage[normalem]{ulem}
\usepackage{subfigure}
\usepackage[breaklinks, plainpages=false, colorlinks=true, anchorcolor=cyan, linkcolor=red, citecolor=cyan, urlcolor=magenta, bookmarks=false]{hyperref}
\usepackage{natbib}
\begin{document}
\renewcommand{\thefigure}{\arabic{figure}}
\setcounter{figure}{0}
\def\I{{\rm i}}
\def\E{{\rm e}}
\def\D{{\rm d}}

\title{Testing Weyl Gravity at Galactic and Extra-galactic Scales}

\author{Koushik Dutta}
\email[]{koushik.physics@gmail.com}
\affiliation{ Theory Divison, Saha Institute of Nuclear Physics, HBNI,1/AF Bidhannagar, Kolkata- 700064, India}

\author{Tousif Islam}
\email[]{ti13ms011@iiserkol.ac.in}
\affiliation{ Department of Physical Sciences, Indian Institute of Science Education and Research Kolkata, Mohanpur- 741246, India}
\affiliation{ International Centre for Theoretical Sciences, Tata Institute of Fundamental Research, Bangalore- 560012, India}

\begin{abstract}
We examine the viability of Weyl conformal gravity as an alternative to the general theory of relativity. By using the extended rotation curve of the Milky Way and velocity dispersions of four globular clusters, we show that Weyl gravity predictions without resorting to dark matter comply with observations at the galactic scale. For the Milky Way, we demonstrate that the uncertainty in baryonic modelling results a bracket of possible rotational velocities which well encompasses the diversity in rotation curve construction. Such diversity generally arises from differences in measurements of velocity anisotropy parameter, and the circular speed and Galactocentric distance of the Sun. Furthermore, we explore the ability of Weyl gravity to account for the inferred acceleration of Abell cluster 1689.
\end{abstract}
\maketitle
\section{Introduction}
The validity of Einstein's General Relativity (GR) is well-established in solar-system neighborhood and binary pulsar systems \cite{solar1}. The recent detection of gravitational wave by LIGO \cite{gw} has further extended its credibility to dynamical strong gravity regimes. However, the theory is plagued by an apparent `mass-discrepancy' in galaxies and clusters. These discrepancies have motivated the ad-hoc addition of mysterious `Dark Matter' (DM) in the current cosmological paradigm which considers GR to be valid at all length-scale. However, ambitious experiments designed to detect dark matter have so far failed to give any positive result \cite{bertone2005particle}.\\

Alternatively, the `mass-discrepancy' could be interpreted as the manifestation of new gravitational physics. The nature of gravity might be intrinsically different at galactic and cosmological scales. This idea encouraged the emergence of a number of modified or alternative theories of gravity. One of the most popular alternative gravity models is Weyl conformal gravity (CG). The theory has recently gained momentum because of its grounding in field theory, embedded local invariance principle and interesting cosmology with naturally arising inflation \cite{weylrot5}. The promises of fourth order terms in Weyl gravity to prevent the Big Bang singularity of GR \cite{weylmotivation1} and to be one-loop re-normalized \cite{weylmotivation2} has created further interest. Moreover, Mannheim and O'Brien have successfully explained the observed galactic rotation curves for a number of galaxies using Weyl gravity without invoking dark matter \cite{weylrot1,weylrot2,weylrot3,weylrot4}. Subsequent studies have confirmed that rotation curve analysis in Weyl gravity is consistent with perihelion precession of mercury \cite{perihelion} and bending of light issues \cite{bending1,sultana,cattani}.\\

One of the predictions of Weyl gravity is the eventual decline of galactic rotational curves \cite{weylrot3}. Galaxies with observational data for rotational velocity profiles extending way beyond the optical length could therefore be utilized to test Weyl gravity. Over the last decade, observed Milky Way (MW) rotation curve has been obtained starting from its innermost regions out to distances beyond 100 kpc from the galactic center using kinematical data of a variety of tracer objects [ Sofue et al (YS09) \cite{sofue1}; Xue et al (X08) \cite{xue}; Sofue (YS12) \cite{sofue2};  Bhattacharjee et al (BCK14) \cite{pijush}; Huang et al (YH15) \cite{yh16} ]. However, the construction of the Milky Way rotation curve heavily relies on three galactic parameters: galacto-centric distance $R_0$ and circular velocity of the Sun $V_0$ and anisotropy parameter $\beta$. Till date, these three fundamental parameters remain remarkably uncertain. O'Brien and Moss (OM15) \cite{obrien} has recently compiled the rotational velocity data from YS09, X08 and BCK14 and fitted within the context of Weyl gravity with mass-to-light ratio as the only free parameter in the model. Though they have found an acceptable mass-to-light ratio, it is to be noted that the differences between astronomical datasets are often systematic. A straight-forward fitting to the combined data set from different surveys could therefore potentially over or under-estimate the total mass in the Milky Way. Thus, a more stringent test for Weyl gravity with extended MW rotation curve is due.\\

Another intriguing set of testing grounds for modified gravity theories is the galactic globular clusters (GCs). The projected radial velocity dispersion for several GCs has been found to be maximum at the center and then to eventually decline towards an asymptotic constant value at large radii \cite{ngc1851a1904,ngc288,ngc5139,ngcothers,ngcothers2}. However, GCs are generally believed to contain little or no dark matter \cite{gcdm1,gcdm2,gcdm3}. Therefore, the velocity dispersion has been expected to follow a Keplerian fall-off and ultimately vanish at larger radii if GR (and Newtonian gravity, weak field limit of GR) would have been valid at GCs. Although classical phenomenon like tidal heating could have been a possible Newtonian gravity explanation for the apparent increase of velocity dispersion in the outskirts of GCs, no solid support for such hypothesis has been found \cite{gctidal}. On the other hand, the eventual flattening of velocity dispersions in GCs hints an interesting analogy with flat rotation curves in elliptical galaxies. Therefore, it might be more logical to argue that the flatting out of velocity dispersions in different GCs have a common origin and is linked to the breakdown of GR at those scale.\\

At this point, we identify a third front to test Weyl gravity predictions. Recently, acceleration profile of Abell cluster 1689 has been inferred \cite{nieu1} from lensing data. It has been claimed that popular alternative gravity theories like Modified Newtonian Dynamics (MOND) \cite{mond1} and Moffat's MOdified Gravity (MOG) or scalar-tensor-vector-gravity \cite{mog} cannot fit the acceleration profile unless an additional dark matter profile ( such as heavy neutrinos ) is assumed. The inferred acceleration profile of A1689 thus provides a crucial extra-galactic test for Weyl gravity.\\

This article aims to explore the astrophysical viability of Weyl gravity. Our work expands from galactic scale up to the length-scale of clusters. First, we test Weyl gravity against the Milky Way rotation curve data. Our approach differs significantly from OM15 \cite{obrien}. We intend neither to compile rotational data from different surveys nor to fit any of them. Rather, we adopt a state-of-art mass model from \cite{mcmillan,bulge,bulge2} and predict the mean rotation profile for the Milky Way up to around 120 kpc and then compare it with observed rotational velocity curve reported by BCK14 \cite{pijush}. The reason for choosing the data set from BCK14 \cite{pijush} is that the assumed values for the galactic constants [$R_0$,$V_0$] in their study closely matches with the most up-to-date measurements from VERA and VLBA surveys \cite{reid}. Furthermore, we show that the embedded uncertainties in the mass model results a `bracket' of rotational velocities possible in the Milky Way within the context of Weyl gravity. Whether this baryon bracketing of rotation curve can successfully encompass the variation in observational data \cite{sofue2,pijush,yh16}, which arises due to the uncertainty in velocity anisotropy parameter and circular velocity at the solar position, is a prime focus of our study. This analysis is done in Section \ref{sec3}. In subsequent section \ref{sec4}, we extend our analysis to globular clusters. We choose a set of four GCs whose distance (from galaxy center), luminosities and sizes are very different from each other. Therefore, we expect that if there is any systematic in their velocity dispersion which hints a Newtonian breakdown, Weyl gravity would be able to capture that. In Section \ref{sec5}, we construct the baryonic mass profile of A1689 with parameterized models for the galaxies \cite{abel1} and inter-cluster gas \cite{nieu2} and compute the Weyl gravity acceleration for the cluster. The predicted acceleration profile is then compared with the one inferred from lensing surveys. Finally, in Section \ref{sec6}, we discuss several aspects of our results and draw conclusions.
\section{Weyl Conformal Gravity}
Though conformal theory of gravity was originally developed by Weyl, the theory has later been re-studied by Mannheim and Kazanas \cite{weylrot5,weyl1}. In addition to the coordinate invariance, Weyl gravity employs the principle of local conformal invariance of the space-time geometry
\begin{eqnarray}
g_{\mu \nu} (x) \rightarrow \Omega^{2}(x) g_{\mu \nu} (x),
\label{eq1}
\end{eqnarray}
where $\Omega(x)$ is a smooth strictly positive function. Imposition of such requirement leads to the unique action in Weyl gravity
\begin{align}
&
\begin{aligned}[t]
&I_{w}\\
&=-\alpha_{g} \int d^{4}x \sqrt{-g} C_{\lambda\mu\nu\kappa} C^{\lambda\mu\nu\kappa}  \\
&= -2 \alpha_{g} \int d^{4}x \sqrt{-g} [ R_{\lambda\mu\nu\kappa} R^{\lambda\mu\nu\kappa}- 2 R_{\mu\kappa}R^{\mu\kappa} + \frac{(R^{\nu}_{\nu})^{2}}{3} ] ,
\end{aligned}
\label{eqn2}
\end{align}
where $\alpha_{g}$ is a dimensionless coupling constant and $C_{\lambda\mu\nu\kappa}$ is the Weyl tensor which is expressed as a combination of the Riemann tensors, Ricci tensors and the Ricci scalar :
\begin{align}
& 
\begin{aligned}[t]
&C_{\lambda\mu\nu\kappa}\\
&=R_{\lambda\mu\nu\kappa} - \frac{1}{2}(g_{\lambda\nu}R_{\mu\kappa} 
- g_{\lambda\kappa}R_{\mu\nu} - g_{\mu\nu}R_{\lambda\kappa} + g_{\mu\kappa}R_{\lambda\nu} ) \\
&+ \frac{1}{6}   R^{\alpha}_{\alpha}( g_{\lambda\nu}g_{\mu\kappa} - g_{\lambda\kappa}g_{\mu\nu} ).
\end{aligned}
\label{eqn3}
\end{align}
Conformal symmetry excludes the conventional Einstein-Hilbert term and therefore does not provide any limit at which Weyl gravity could reduce to   the  standard GR. The symmetry also forbids the presence of any cosmological constant and thus naturally addresses one of the notorious problems in GR \cite{weylrot5}.\\

A  functional variation of the Weyl action with respect to the metric $g_{\mu\nu}$ results the following fourth order gravitational field equation in Weyl gravity:
\begin{eqnarray}
4 \alpha_{g} W^{\mu\nu} = 4 \alpha_{g} (C_{;\lambda;\kappa}^{\lambda\mu\nu\kappa} - \frac{1}{2} R_{\lambda\kappa} C^{\lambda\mu\nu\kappa} ) = T^{\mu\nu} , 
\label{eqn4}
\end{eqnarray}
where $T^{\mu\nu}$ is the matter-energy tensor and   `;' denotes covariant derivative. Since $W^{\mu\nu}$ vanishes when $R^{\mu\nu}$ is zero, a vacuum solution of the field equation in GR is automatically a vacuum solution of Weyl gravity. Thus, Schwarzschild solution is indeed an exact vacuum solution of Weyl gravity. However,   $W^{\mu\nu}=0$  does not necessarily mean $R^{\mu\nu}$ is zero. The highly non-linear character of the field equation makes it difficult to obtain any analytical solution. However, Mannheim and Kazanas have been able to find an exact vacuum solution in the case of a static, spherically symmetric geometry \cite{weyl1,weylrot5} with the line element
\begin{equation}
ds^2= \left[-B(r) dt^2+ {dr^2\over B(r)} + r^2
d\Omega_2\right]~~,
\label{20}
\end{equation}
where the metric coefficient is given by
\begin{eqnarray}
B(r)=  1 - \frac{2\beta}{r} +\gamma r -kr^{2},
\label{23}
\end{eqnarray}
with $\beta$,$\gamma$ and $k$ being constants. \\

In  order to determine the gravitational potentials of realistic sources, it is necessary to obtain solutions associated with sources in   the  weak field limit.   It  could be shown that, for spherically symmetric sources, the non-linear field Eq. (\ref{eqn4}) dramatically reduces to   a  remarkably simple fourth order Poisson equation \cite{weylrot5}:
\begin{equation}                                                                               
\nabla^4 B(r) = \frac{3}{4\alpha_g B(r)}\left(T^0_{{\phantom 0} 0} -
T^r_{{\phantom r} r}\right) =  f(r) ~~.
\label{45}
\end{equation}
\noindent The function $f(r)$ represents a convenient source function whose form is not fixed \textit{a priori}. The Newtonian limit or non-relativistic weak field limit only changes the source function. In case of a perfect fluid, $T_{00}= \rho(r) B(r)$ and $T_{rr}= p(r)/B(r)$,
where $p(r)$ and $\rho(r)$ are the pressure and energy density respectively. Therefore $T^{0}_{0}-T^{r}_{r}= - \rho(r) - p(r)$. For slowly moving sources, $p(r) \approx 0$ and thus $T^{0}_{0}-T^{r}_{r} \approx - \rho(r)$ where $\rho(r)$ is the mass density. Therefore, the source function becomes $f(r)\approx - \rho(r)$. The general solution of Eq.  (\ref{45})  could readily be obtained employing Greens' function method \cite{weylrot5}  :  
\begin{align}
B(r)  &= 
\begin{aligned}[t]
&-\frac{r}{2} \int^{r}_{0}dr^{'}r^{'2}f(r^{'}) - \frac{1}{6r} \int^{r}_{0}dr^{'}r^{'4}f(r^{'})  \\
&-\frac{1}{2} \int^{\infty}_{r}dr^{'}r^{'3}f(r^{'}) -\frac{r^{2}}{6} \int^{\infty}_{r}dr^{'}r^{'}f(r^{'}) + B_{h}(r) ,
\end{aligned}
\label{46}
\end{align}
where $B_{h}(r)$ obeys $ \bigtriangledown^{4}B_{h}(r) = 0$. While the first two integrals originate from the matter distribution inside the source, the third and fourth integrals are attributed to the global matter distribution exterior to it. Hence, a correct study of rotational motions within galaxies and clusters should include both the local contributions from luminous sources in the galaxies/clusters and a global contribution of mass outside the galaxy/cluster. A comparison   between Eq. (\ref{46}) and Eq. (\ref{23})  helps to identify:
$\gamma = $ $-\frac{1}{2}$ $\int^{r}_{0}dr^{'}r^{'2}f(r^{'})$ ; $2\beta= \frac{1}{6} \int^{r}_{0}dr^{'}r^{'4}f(r^{'})$; and $k = \frac{r^{2}}{6} \int^{\infty}_{r}dr^{'}r^{'}f(r^{'})$. It could thus be concluded that $\beta$ and $\gamma$ originates completely from the local mass distribution , and  $k$ has a global origin. Identifying $B(r)=1+2\phi/c^{2}$, the gravitational potential for a point source in static, spherically symmetric case is thus: $\frac{\phi(r)}{c^{2}}= \frac{\beta}{r} + \frac{\gamma r}{2}$. Therefore, a star with mass $M$, radius $r_0$ and   normalized source function $f^{*}(r)=f^{}(r)/M_{\odot}$  will yield a normalized gravitational potential featuring a Newtonian term as well as a linear one:
\begin{eqnarray}
V^{\ast}_{source}(r > r_{0}) = - \frac{\beta^{\ast}c^{2}}{r} + \frac{\gamma^{\ast}c^{2}r}{2},
\end{eqnarray}
where $\beta^{\ast}=(\frac{M_\odot}{M})\beta$ and $\gamma^{\ast}=(\frac{M_\odot}{M})\gamma$. The global effects in Weyl gravity have two different origins: a homogeneous and isotropic cosmological background, and large scale inhomogeneities in the forms of galaxies, clusters and filaments. The gravitational potential due to homogeneous cosmological background could be expressed as $V_{homo} = \frac{\gamma_0 c^{2} r}{2}$ \cite{weylrot2,weylrot5}. On the other hand, the inhomogeneities in the cosmological background will contribute through the third and fourth integral of Eq. (\ref{46}) and is found to be $V_{inhom} = - \kappa c^{2} r^{2}$ (where $\kappa$ is a constant). Therefore, total gravitational potential due to   the  global distribution of matter is $V_{global} = \frac{\gamma_0 c^{2} r}{2} - \kappa c^{2} r^{2}$.
\section{Testing Weyl Gravity with the Milky Way Rotation Curve}
\label{sec3}
\subsection{Milky Way Mass Model}
\label{sec31}
\begin{figure}[h]
	\begin{center}
		\includegraphics[scale=0.4]{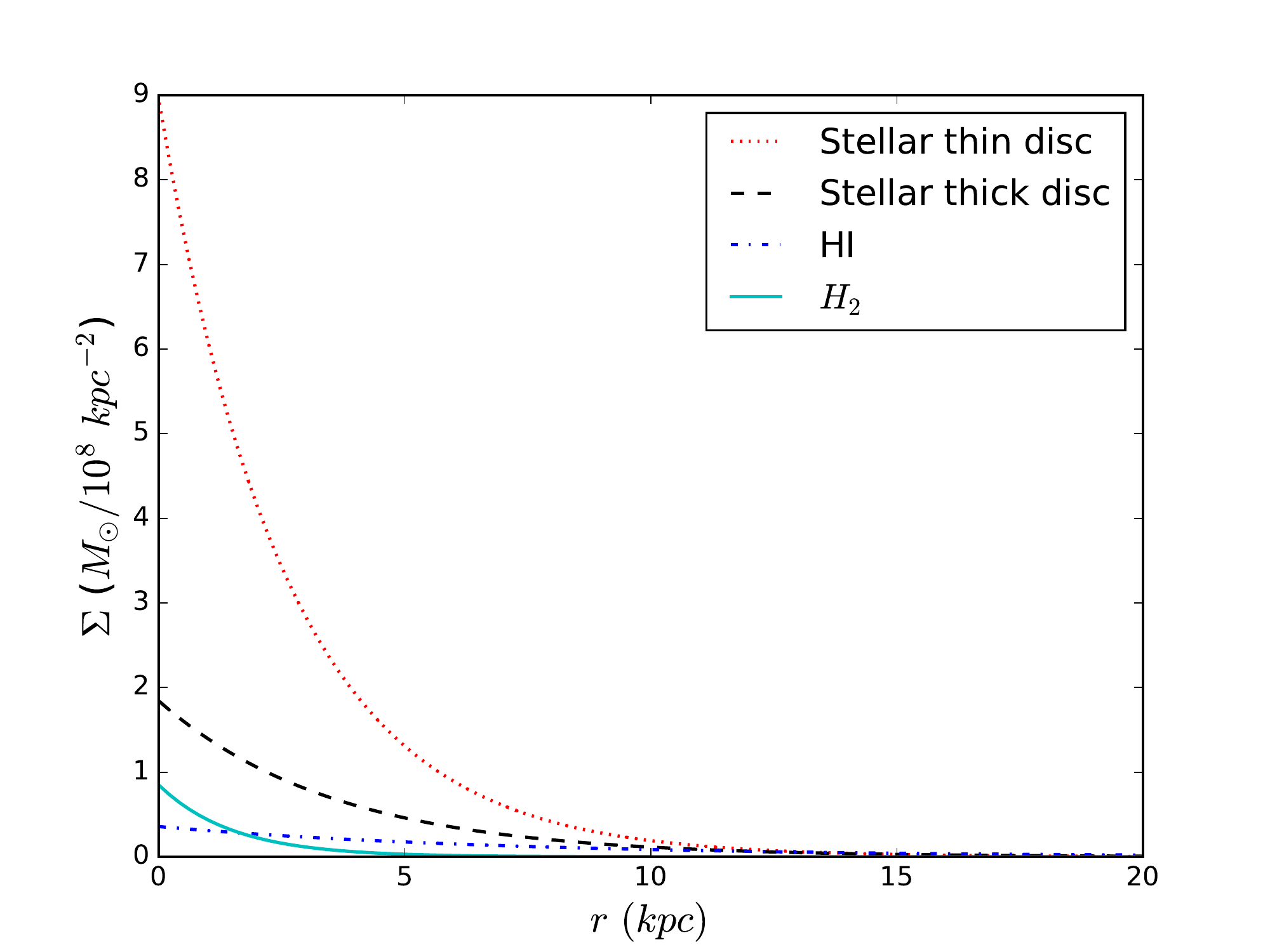} 
	\end{center} 
	\caption{\textbf{Plot shows the surface mass density of different disk components of Milky Way up to r$<$20 kpc. Stellar mass   dominates the gas mass   in this region.}}
	\label{fig1}
\end{figure}

In this study, we use a simple but detailed mass model of the Milky Way. We decompose the Milky Way into five distinct axis-symmetric components: a spherical central bulge, thin and thick stellar disks, and HI and molecular gas disks.   The parts of the galaxy we consider in this rotation curve analysis lie beyond any central `hole' and thus the presence of holes does not affect our results at all. Therefore, we do not include any disk `hole' in our model.   Following McMillan \cite{mcmillan} and Mannheim \cite{weylrot5}, we use exponential mass profile with varying scale length for each of the disk components: $\Sigma_{i}(r)=\Sigma^{0}_{i} e^{-r/R_{i}}$   ($i$ refers to individual disk components)   , where   $\Sigma$, $\Sigma^{0}$ and $R$ denotes the surface mass density, maximum surface density (at the center) and scale length for respective components   respectively   (Figure \ref{fig1}). Disk mass enclosed within the distance $R$ could easily be calculated as : $M_{i}=2 \pi \Sigma^{0}_{i} R^{2}_{i}$. The values for different parameters has been taken from McMillan \cite{mcmillan} and   are listed in Table \ref{T1}.  \\

\noindent For the stellar bulge, we adopt a more convenient exponential density considered in \cite{bulge}
\begin{equation}
I(R)=\frac{N}{2\pi t^2}e^{-R/t},
\label{B5}
\end{equation}
which yields a three dimensional mass density
\begin{equation}
\sigma(r)=\frac{N}{2\pi^2 t^3}K_0(r/t),
\label{B6}
\end{equation}
where $N$ is the total number of solar mass stars in the bulge and $t$ is the extent of the bulge. We have used $M_{bulge} = 2.0 \pm 0.3 \times 10^{10} M_{\odot}$ \cite{bulge2} and $t=1$   kpc  (following \cite{weylrot1}). The very definition and measurements of radial extent varies in different study. However, we identify that such studies report a scale within a range of   0.6 kpc to 2.0 kpc   and thus we decide to stick to a crude but average estimate of the bulge length $t=1$ kpc. The overall number of the stars in the spherical bulge could be calculated via $N=\frac{M}{M_{\odot}}$. Finally, we have included a central super-massive black hole with a mass $M_{bh} = 4.0\pm0.3\times10^{6} M_{\odot}$ in our model.\\

\begin{table}
	\centering
	\caption[Milky Way mass model]{\textbf{Parameters for the Milky Way mass model \cite{mcmillan}}}
	\label{T1}
	\begin{tabular}{l    l}
		\hline \\
		$\Sigma^{0}_{thin}$ & $ 886.7 \pm 116.2$ $ M_{\odot} pc^{-2}$ \\
		$R_{thin}$ & 2.6 $\pm$ 0.52 kpc\\ 
		$\Sigma^{0}_{thick}$ & $ 156.7 \pm 58.9$ $ M_{\odot} pc^{-2}$ \\
		$R_{thick}$ & 3.6  $\pm$ 0.72 kpc\\ 
		$M_{HI}$ & $ 1.1 \times 10^{10} M_{\odot}$ \\
		$R_{HI}$ & 7.0 kpc\\
		$M_{H2}$ & $ 1.2 \times 10^{9} M_{\odot}$ \\
		$R_{H2}$ & 1.5 kpc\\
		& \\
		\hline 
	\end{tabular} 
\end{table}
\subsection{Weyl gravity Prediction}
\label{wp}
We model   each   disk components of galaxies with a typical exponential surface mass distribution $\Sigma(r)=\Sigma_0 e^{-r/R_0}$ where $R_{0}=1/ \alpha$ is the scale length and $N=2\pi\Sigma_0 R^{2}_{0}$ is the total number of stars \cite{weylrot5,weylrot2,weylrot4}.    Each   star in the disk generates a potential $V^{\ast}_{star}(r > r_{0}) = - \frac{\beta^{\ast}c^{2}}{r} + \frac{\gamma^{\ast}c^{2}r}{2}$. The resultant potential due to a disk component of the galaxy could thus be obtained by an integration over the entire disk. The total contribution to rotational velocities from the luminous mass within the disk is found to be \cite{weylrot5}
\begin{align}
& 
\begin{aligned}[t]
&v^{2}_{disk}(r)\\
&=\frac{N\beta^{\ast}c^{2}r^{2}}{2R^{3}_{0}} \Big[ I_{0}\left(\frac{r}{2R_{0}}\right)K_{0}\left(\frac{r}{2R_{0}}\right) -I_{1}\left(\frac{r}{2R_{0}}\right)K_{1}\left(\frac{r}{2R_{0}}\right) \Big]\\
& + \frac{N\gamma^{\ast}c^{2}r^{2}}{2R_{0}} I_{1}\left(\frac{r}{2R_{0}}\right)K_{1}\left(\frac{r}{2R_{0}}\right),
\end{aligned}
\label{m4}
\end{align}
where $I_0$, $I_1$, $K_0$ and $K_1$ are modified Bessel functions. While the first term in Eq.  (\ref{m4})   is the contribution from the Newtonian term $\beta$, the second term originates from the linear potential. On the other hand, spherical bulge with mass profile similar to the one in Eq.  (\ref{B6})   yield circular velocities of the form \cite{weylrot5}
\begin{align}
& 
\begin{aligned}[t]
&v^2_{bulge}(r)\\
&={N\gamma^* c^2 r\over \pi} \int_0^{r/t}dz\, z^2K_0(z)\\
&- { N\gamma^* c^2 t^2 \over 3 \pi r} \int_0^{r/t}dz\, z^4K_0(z)+{2 N\gamma^* c^2 r^3 \over 3\pi t^2}K_1(r/t).
\end{aligned}
\label{m7}
\end{align}
One thus obtains the rotational velocity prediction for the Milky Way galaxy due to the local mass distribution as
\begin{align}
v^2_{loc}(r)=&
\begin{aligned}[t]
&v^2_{bulge}(r) + v^2_{disk,thin}(r) + v^2_{disk,thick}(r)\\
&+ v^2_{disk,HI}(r) + v^2_{disk,H2}(r).
\end{aligned}
\label{m8}
\end{align}
Finally, on including the global effects we find the net rotational velocity in   Weyl   gravity
\begin{eqnarray}
v^2_{tot}(r) = v^2_{loc}(r) + \frac{\gamma_0 c^{2} r}{2} - \kappa c^{2} r^{2}.
\label{m9}
\end{eqnarray}
Previous Weyl gravity fit to rotation curves of 111 galaxy samples by Mannheim and O'Brien \cite{weylrot1,weylrot2,weylrot3} yielded the following best-fit values for the four universal Weyl gravity parameters: $\beta^{\ast} = 1.48 \times 10^{5}$ $cm$; $\gamma^{\ast} = 5.42 \times 10^{-41}$ $cm^{-1}$; $\gamma_0 = 3.06 \times 10^{-30}$ $cm^{-1}$ and $\kappa = 9.54 \times 10^{-54} $$cm^{-2}$. In this study, we will use this set of parameter values to compute the predicted velocity (or acceleration) profile in Weyl gravity. 
\subsection{Rotation Curve data}
BCK14 \cite{pijush} have constructed high quality rotation curve  of  the Milky Way starting from its very inner regions (few hundred pc) out to a large galactocentric distance beyond $\sim$ 100 kpc using kinematical data of different tracer objects moving in the gravitational potential of the Galaxy, without assuming any theoretical models of the visible and dark matter components of the Galaxy. The circular velocities and their respective errors for each of the disk tracer samples have been calculated directly from its known radial distance and measured line-of-sight velocity. On the other hand, for non-disk tracers, rotational velocity has been extracted using Jeans equation which relates the number density and their galactocentric  radial velocity dispersion. It has been found that the mean rotational velocities steadily decreases beyond 60 kpc.
\subsection{Results}
\begin{figure}[h!]
	\begin{center}
		\includegraphics[scale=0.45]{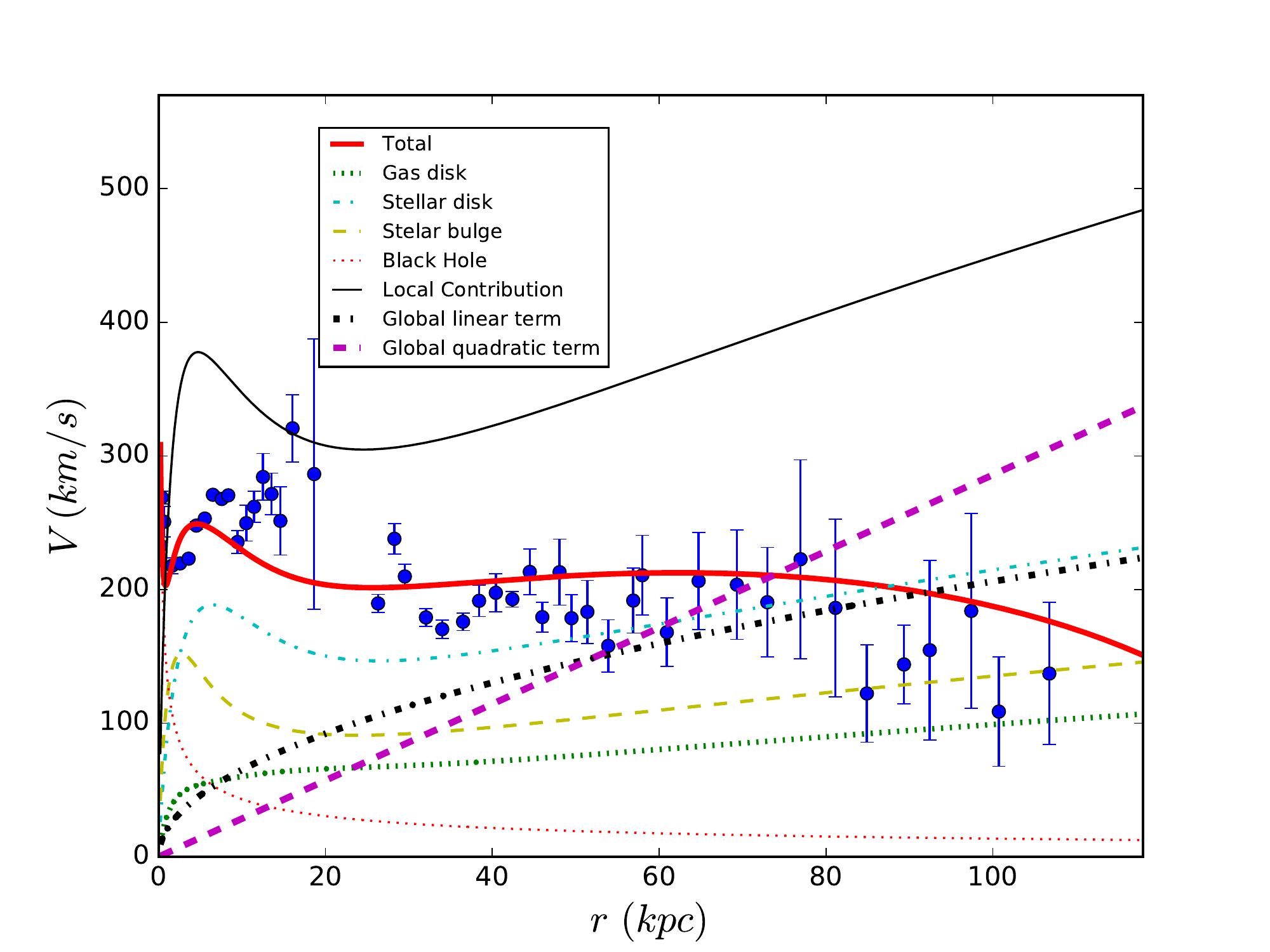} 
	\end{center} 
	\caption{\textbf{Plot shows Weyl gravity prediction (thick red) along with data from BCK14 \cite{pijush}. Contribution from different mass components have been shown separately: stellar disk (cyan dash dotted), gas disk (green thick dotted), stellar bulge (yellow double dashed), Black-hole (red thin dotted). Total local contribution has been plotted in black (thin, lined) while global contributions are shown in thick dashed (due to quadratic term) and thick dash dotted (due to linear term) curves.}}
	\label{fig3}
\end{figure}
\subsubsection{Weyl Gravity: prediction vs observation}
We present the mean predicted rotational velocity profile in Weyl gravity  (using the parameter values mentioned at the end of Section \ref{wp})  for the Milky Way along with   the  data from BCK14. To begin with, we have particularly chosen the rotation curve data constructed with galactic constant [$R_0$,$V_0$] = [8.3 kpc, 244 km/s]. This particular choice of galactic constants is more consistent with recent observations of masers and stellar orbits around SgrA*, the central super-massive blackhole in our galaxy \cite{mcmillan2010uncertainty}. Figure \ref{fig3} shows that Weyl gravity prediction is in   reasonable  agreement with rotation curve data. The predicted rotation curve remains almost flat from 30 kpc to 70 kpc beyond which it gradually falls. In Weyl gravity, a   competing effect  between local and global contributions results the ultimate velocity curve. In Figure \ref{fig3}, we   also  show the contributions from local source mass distribution as well as global effects separately. While local effects dominate within 30 kpc, global effects become the deciding factor beyond 60 kpc. This results an immediate fall-off. Furthermore, the increase in rotational velocities from local contribution due the linear term in the region between 30 kpc and 70 kpc is compensated by a decrease originating from the negative quadratic term associated with the global contributions.

\subsubsection{Effects of the supermassive blackhole}
\begin{figure}[h]
	\begin{center}
		\includegraphics[scale=0.45]{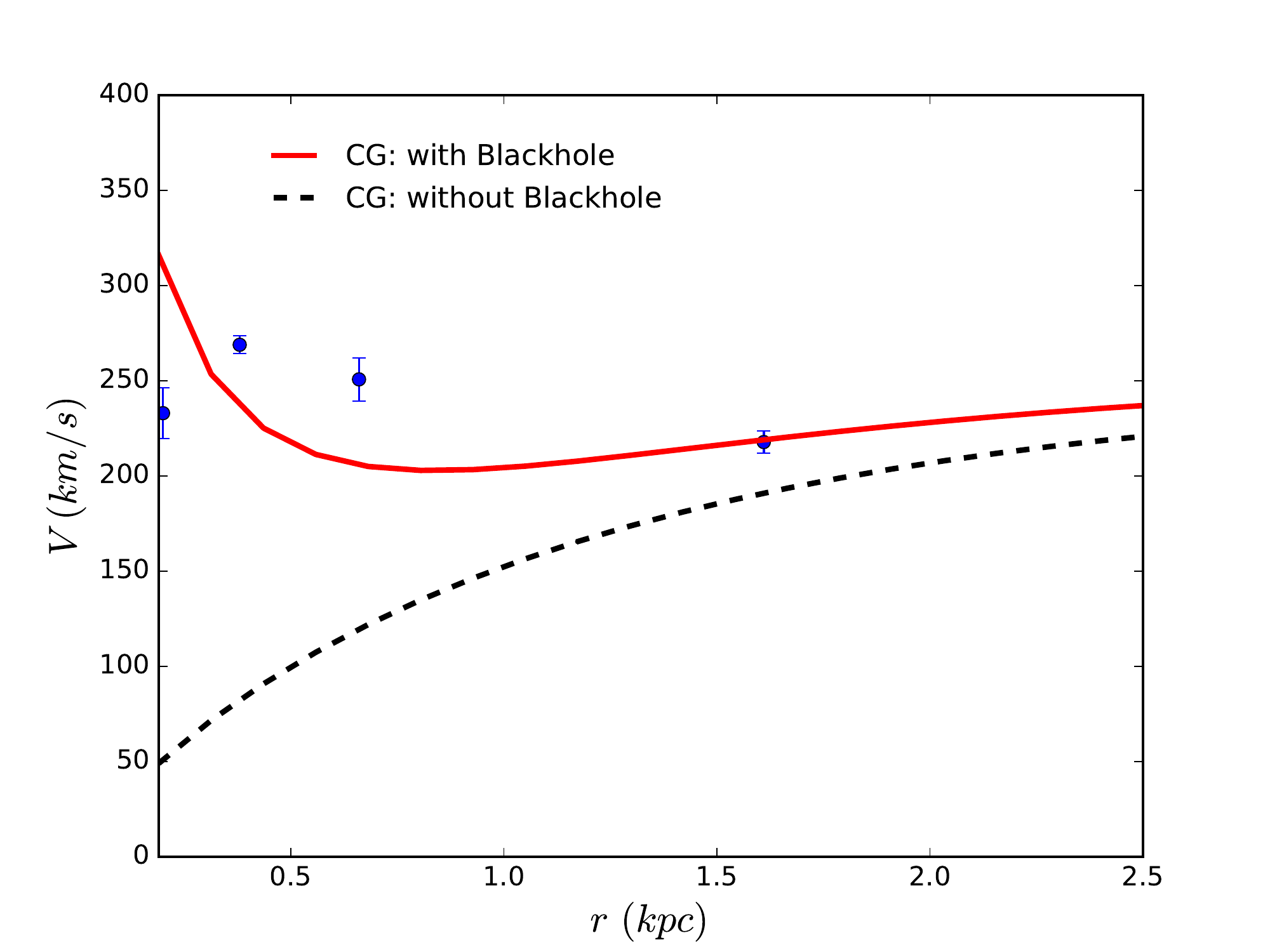} 
	\end{center} 
	\caption{\textbf{Innermost region: inclusion of a supermassive Black-hole in the mass model helps better account for the data. Weyl gravity prediction with Blackhole has been plotted in red (lined) while prediction without a Blackhole is shown in black (dotted).}}
	\label{fig4}
\end{figure}

Weyl gravity prediction not only captures the overall decline in the rotation curve beyond 60 kpc, it is also found to be able to describe an apparent dip around 3 kpc. Within the radius of 3 kpc, rotational velocities continue to rise towards the center. This generally hints the existence of a central black-hole in the galaxy. We have already mentioned in section \ref{sec31} that our mass model includes a central super-massive blackhole of mass around $10^{6}$ $M_{\odot}$. This helps us to better match with observation in the innermost region of our galaxy (Figure \ref{fig4}). A mass model without a central black-hole results a huge discrepancy between the predicted and observed galactic rotational velocity. Predicted velocities differ by almost two orders of magnitude. However, the prediction improves dramatically once we consider a supermassive black hole in the model.
\begin{figure*}[htb]
	\centering
	\subfigure[]{\label{fig:s5}
		\includegraphics[scale=0.37]{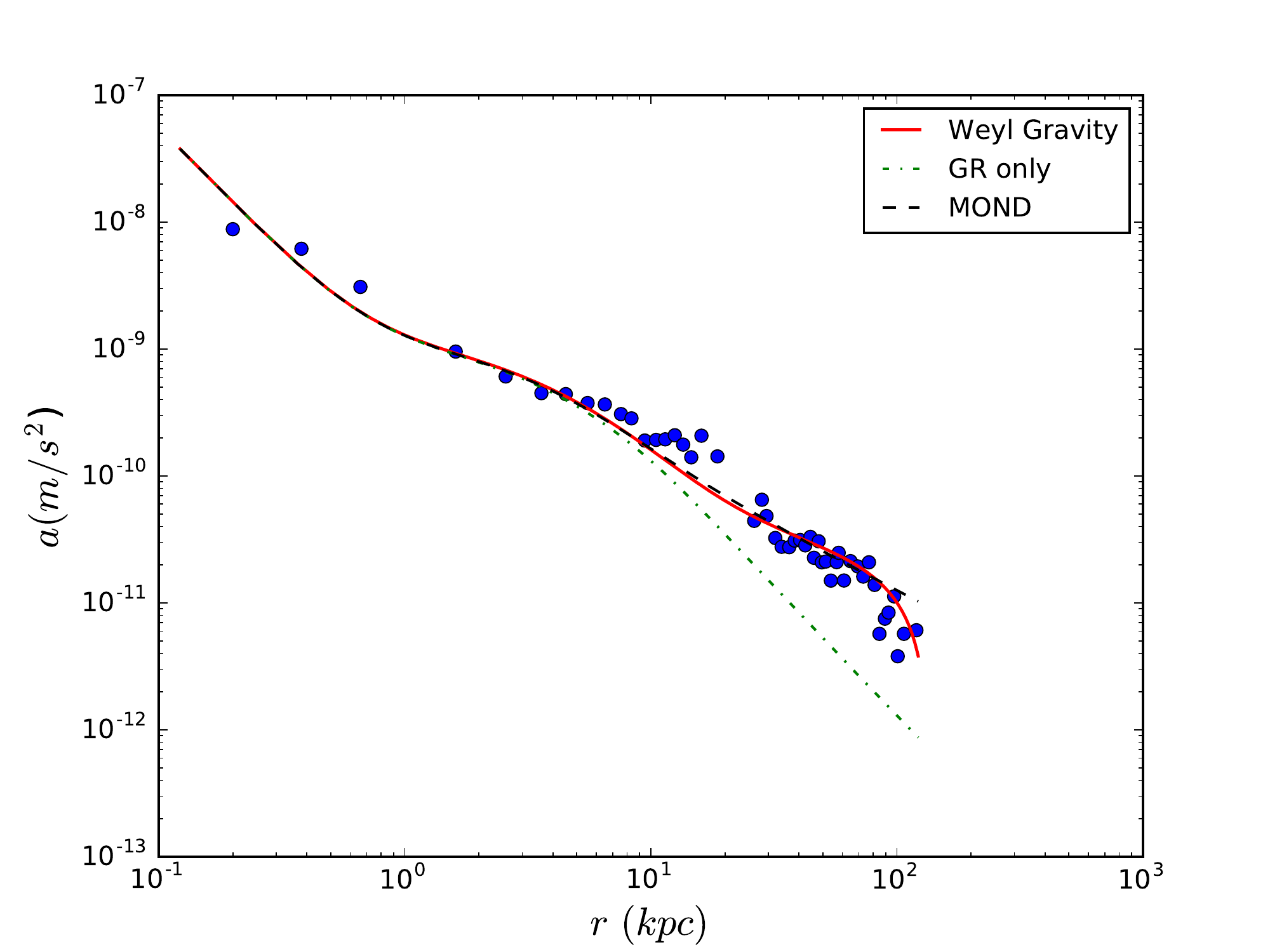}}
	\subfigure[]{\label{fig:s6}
		\includegraphics[scale=0.35]{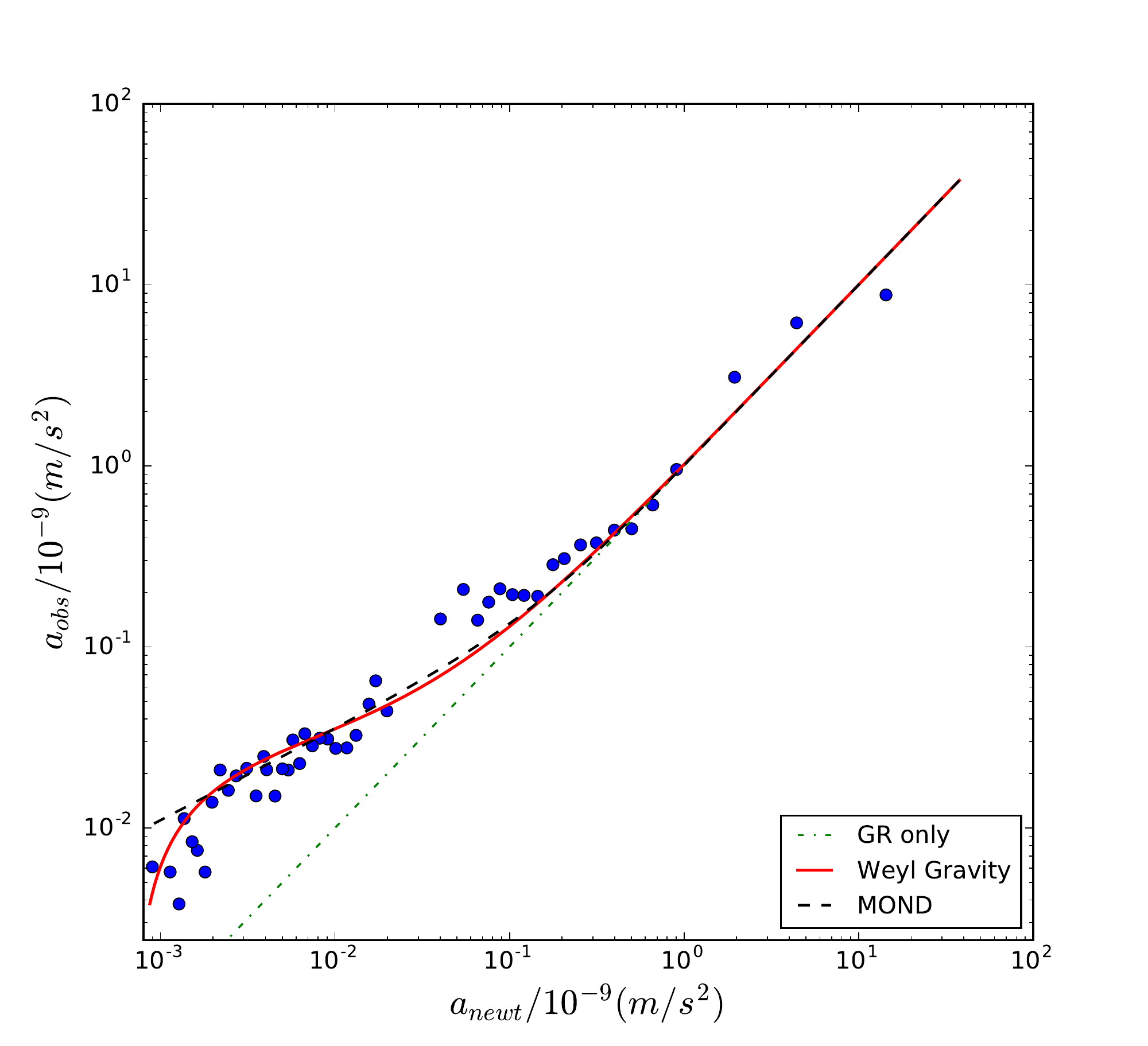}}
	\subfigure[]{\label{fig:s1}
		\includegraphics[scale=0.37]{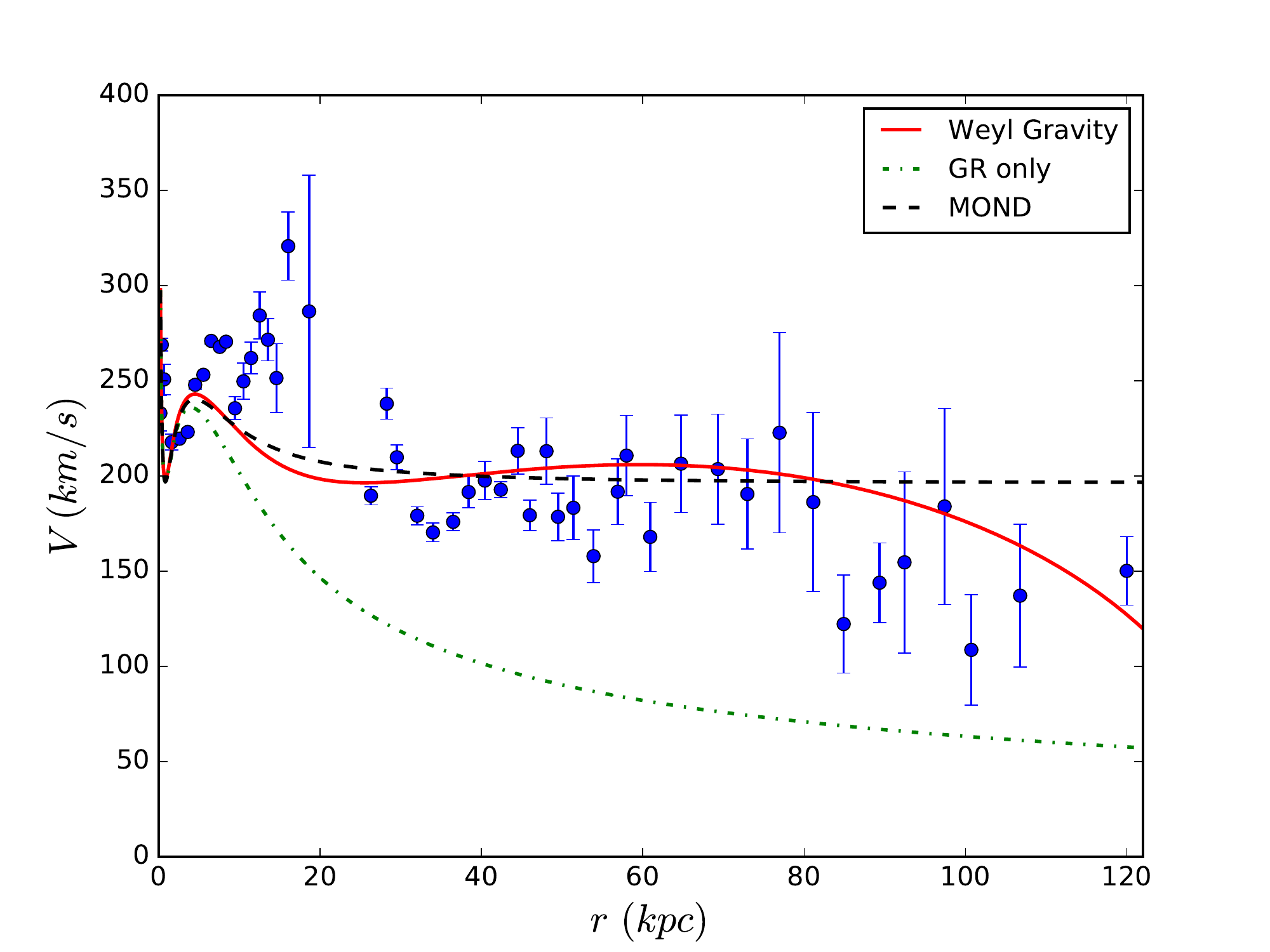}}
	\subfigure[]{\label{fig:s2}
		\includegraphics[scale=0.37]{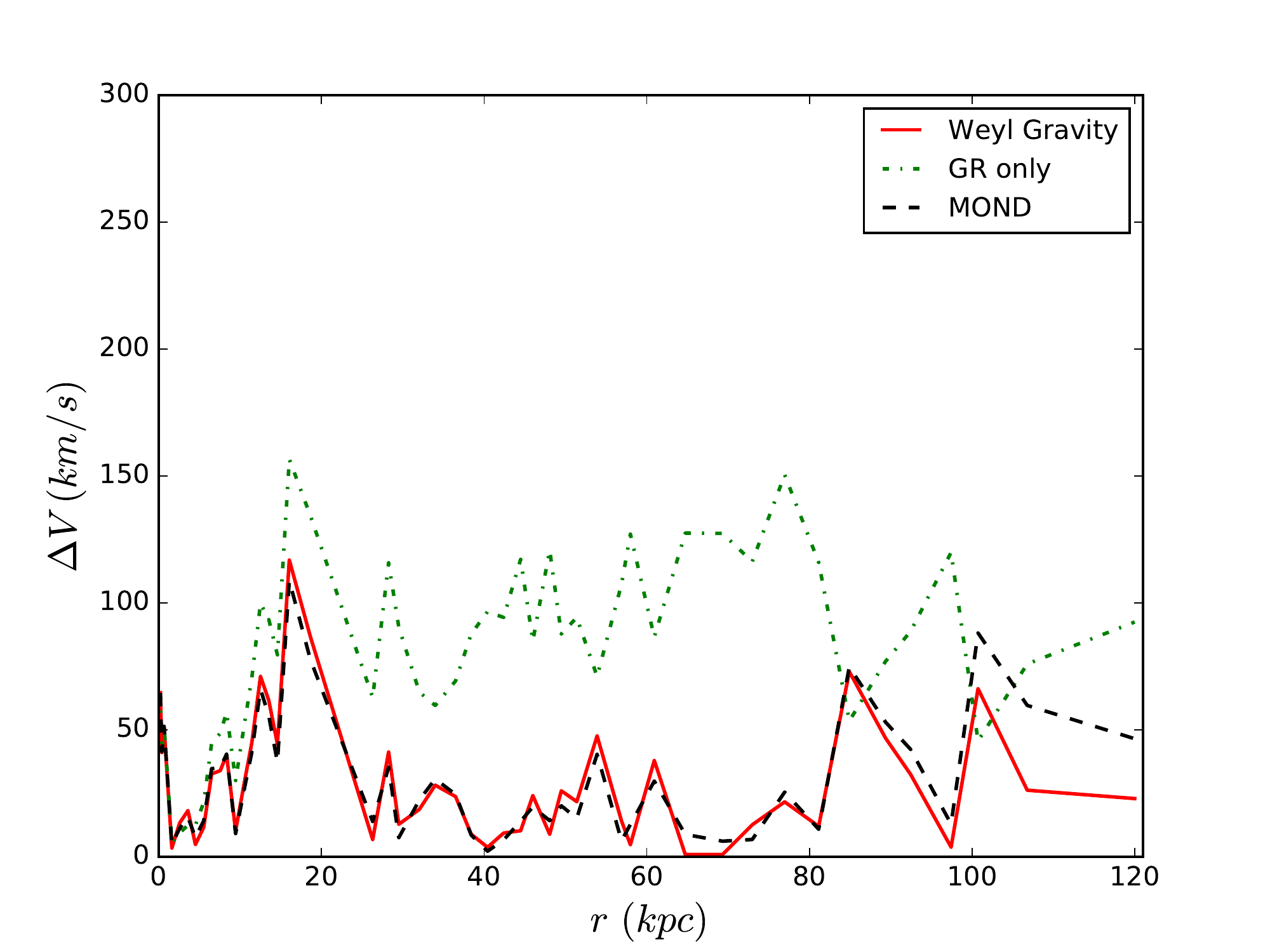}}
	\caption{\label{fig:s} \textbf{(a) loglog plot of observed centripetal acceleration (inferred from BCK14) as a function of radial distances along with predictions in different gravity models ; (b) loglog plot of observed centripetal acceleration as a function of Newtonian expectation. Predicted profiles in different gravity theory as a function of Newtonian expectation are super-imposed. (c) Predicted Rotation curves in different gravity models along with data from BCK14 ; (d) Residual profiles of the rotation curve fit in different gravity models;  [color code: Weyl gravity (red, lined), MOND (black, dashed), GR only (green, dash dotted)]  }}
\end{figure*}

\subsubsection{Radial Acceleration Relation (RAR)}
Recently, McGaugh et al \cite{McGaugh} have established a strong correlation between the observed centripetal acceleration ($a_{obs} = \frac{V^{2}_{obs}}{r}$) and the expected ones ($a_{newt}$) from luminous mass of galaxies. In general, $a_{newt}$ denotes the expected centripetal acceleration in the context of GR (or equivalently in Newtonian gravity) without resorting to dark matter. The advantage of $a_{obs}$ vs $a_{newt}$ plot lies in the fact that both $a_{obs}$ and $a_{newt}$ are independent of each other. While $a_{obs}$ is directly obtained from the observed rotational curve, $a_{newt}$ is generally computed through rigorously solving the Poisson's equation using observed baryonic mass profile as input. There is no guarantee that these two quantity should be correlated if dark matter dominates. Therefore, such strong correlation nullifies the need of dark matter and hints to the modification of  the   laws of gravitation in galactic length scale. This correlation has been found in all types of galaxies irrespective of whether they fall in the low acceleration regime  ($10^{-10}$ $m/s^{2}$ - $10^{-12}$ $m/s^{2}$) or in the high end ($10^{-8}$ $m/s^{2}$ - $10^{-10}$ $m/s^{2}$) . The case for Milky Way is special as its extended rotation curve data spans both the high and low acceleration domain (from $10^{-8}$ $m/s^{2}$ to $10^{-12}$ $m/s^{2}$; Fig \ref{fig:s5} ) and thus offers an unique probe to test any modified gravity model against the radial acceleration relation in the Milky Way. In Fig \ref{fig:s5}, we plot the observed centripetal acceleration as a function of radial distances. We present the radial acceleration relation curve in Fig \ref{fig:s6}. In both cases, Weyl gravity predictions have been superimposed (in red, lined).   The plots reveal that Weyl gravity can very well describe the radial acceleration relation.   
\subsubsection{Comparison with MOND}
In the case of MOND \cite{mond1}, below a critical value of acceleration $a_0$, the acceleration law is   phenomenologically   modified by the introduction of an interpolating function $\mu(x)$ such that
\begin{equation}
	\mu \left(\frac{a}{a_0}\right) a = a_{N}.
\end{equation}
The interpolating function $\mu(x) \approx x$ when $x \ll 1$ and $\mu(x) \approx 1$ when $x \gg 1$. Therefore, at large acceleration Newtonian behavior is recovered. Although the theory can accommodate different forms of interpolating functions, we assume a simple form of $\mu(x=\frac{a}{a_0})$ \cite{mond1}
\begin{equation}
	\mu (x) = \frac{x}{\sqrt{(1 + x^{2})}}
\end{equation}
with $a_0$ = $1.21 \times 10^{-10} m/s^{2}$. This results the following MOND acceleration \cite{mond1}
\begin{equation}
	a_{MOND} =  \frac{a_{N}}{\sqrt{2}}\Big[ 1 +  \Big( 1 + \left(\frac{2a_0}{a_{N}}\right)^2 \Big)^{1/2} \Big]^{1/2},
\end{equation}
where $a_{N}$ is the Newtonian acceleration associated with baryonic masses. The corresponding rotational velocity thus becomes
\begin{equation}
	v^{2}_{MOND}(r)=a_{MOND} \times r.
\end{equation}
\noindent We now compare the ability of   Weyl   gravity to account for the Milky Way rotation curve with popular modified gravity alternative MOND. In Figure \ref{fig:s1}, we plot the predictions of   Weyl gravity (red, lined), MOND (black, dashed) and GR only (without resorting to dark matter) (green, dash dotted)   along with data from BCK14. We find that Weyl gravity and MOND predictions seem to be almost consistent with the data. However, a fall-off in rotation curve beyond 70 kpc is prominent in Weyl gravity. This feature is absent in MOND prediction. In MOND, the predicted rotation curve becomes almost flat for the entire region 30 kpc $< r <$ 120 kpc. It therefore misses to match the last couple of data points. As expected, GR only (i.e. no dark matter) fails to fit the rotation profile alone. However, we must note that when added with a dark matter profile, it should come in agreement with data.\\

In Figure \ref{fig:s2}, we have plotted the residuals of rotation curve fit (predictions) as a function of distances from the galactic center. It could be easily seen that Weyl gravity and MOND produce similar residual profiles. However, residual values for MOND are higher in the outermost region of the galaxy. In order to quantitatively identify the best gravity model, we calculated the reduced chi-square values for the fitting (as a measure for the goodness of fit) and found that Weyl gravity ($\chi^{2}/dof = 7.6$) and MOND ($\chi^{2}/dof = 8.3$) yield almost similar values.  Thus the performance of these two gravity models is comparable. We now plot the predicted centripetal acceleration in MOND and GR as a function of radial distances in log-log scale in Figure \ref{fig:s5}. Though the plot does not hold any new information, it stresses the inability of GR (without dark matter) to comply with the observed acceleration profile. We can easily see that the GR (no dark matter assumed) expectation deviates from observation beyond 10 kpc, where the acceleration falls below $10^{-10} ms^{-2}$. Figure \ref{fig:s6} presents the radial acceleration relation which shows a strong correlation between baryonic mass and observed centripetal acceleration. It has already been noted that  Weyl   gravity can easily account for the radial acceleration relation; but Figure \ref{fig:s6} suggests that MOND does the same and thereby challenges the notion of dark matter. However, a careful analysis shows that MOND overshoots the data in the extreme low end of   the   acceleration.
\begin{figure*}[!ht]
	\includegraphics[scale=0.6]{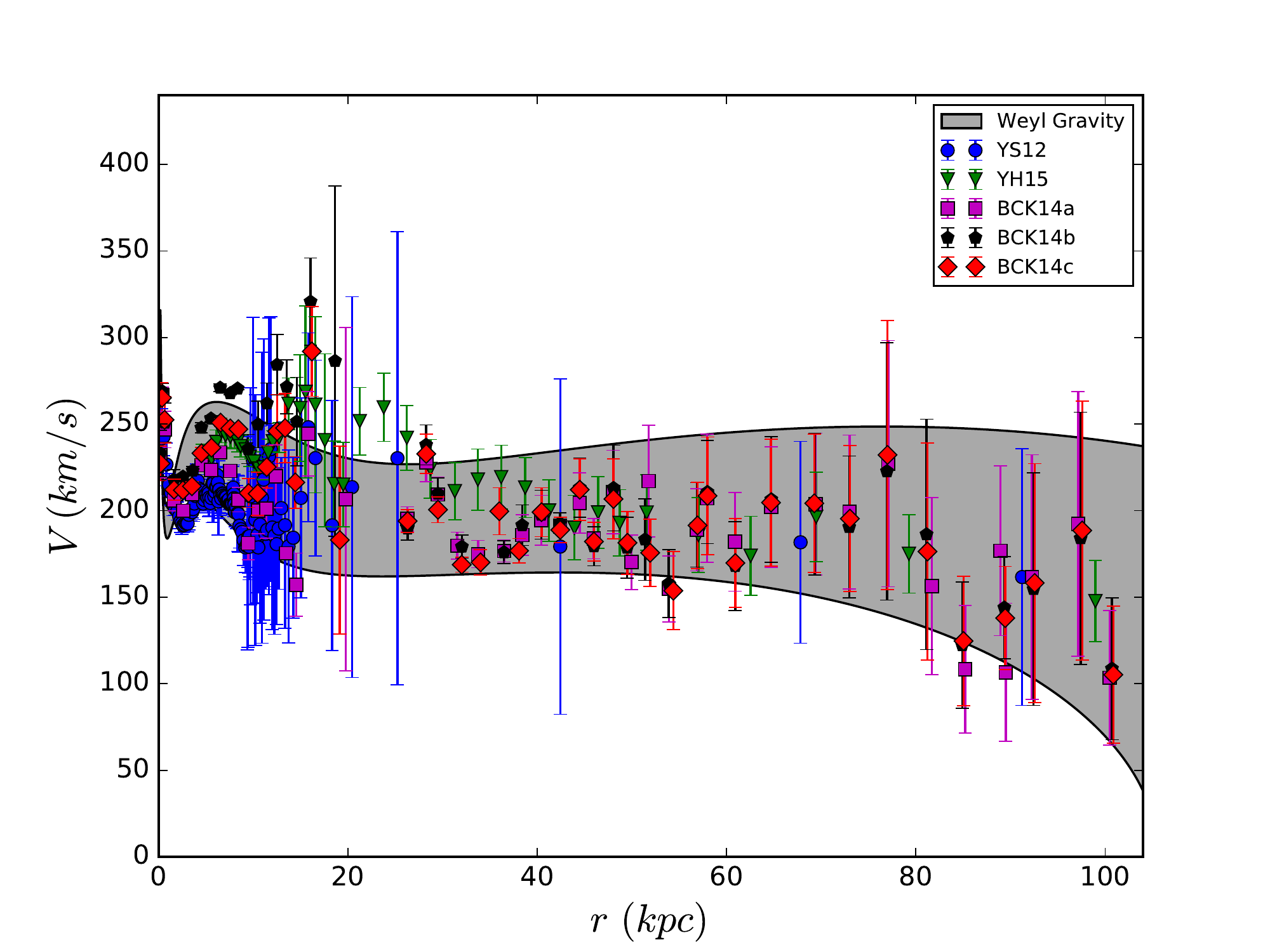} 
	\caption{\textbf{Baryon bracketing in Milky Way in the context of Weyl Gravity. Data from three different groups (BCK14, YS12, YH15) have been super-imposed.}}
	\label{fig6}
\end{figure*}
\subsubsection{Baryon bracket in Weyl Gravity}
Weyl gravity has indeed been found to comply with the Milky Way rotation curve quite well. Still, a more vigil eye would be able to identify a troublesome region: 15 kpc $< r <$ 20 kpc, where the predicted values are significantly lower than observation. While one may set to hunt down possible reasons in the assumed mass model, there is another angle worth exploring. First, we point out that we have tested the Weyl gravity prediction against rotation curve data (BCK14 \cite{pijush}) which has been constructed with galactic constant [$R_0$,$V_0$] = [8.3 kpc, 244 km/s] (BCK14b). While these values of galactic constants are in agreement with a couple of recent studies \cite{reid,mcmillan2010uncertainty}, their values remain remarkably uncertain in   the   literature. Even BCK14 have explored two more sets of galactic constants ([8.5 kpc, 220 km/s] (BCK14c), [8.0 kpc, 200 km/s]) (BCK14a) in their re-construction of rotation profile. Their study assumes a radially varying anisotropy parameter $\beta$. BCK14 have found that rotation curve construction   is highly   sensitive to the choice of $R_0$ and $V_0$ in small radial distances. At larger distances, rotation profiles are hugely dependent on anisotropy parameter. Thus, different choices of $R_0$,  $V_0$ and $\beta$ will lead to variations in the Milky Way rotation curves.\\

Second, to compute Weyl gravity prediction, we chose to calculate the rotation curve only from the mean mass profile. As our mass model usually allows a maximum and minimum values for different disk and bulge parameters, it is definitely possible to compute the maximum and minimum rotation curve profiles in Weyl gravity.   The region between the minimum and maximum rotation curves will represent the allowed values of rotational velocities in Weyl gravity given the baryonic profile of the Milky Way.   Thus, the final piece of analysis would be to compute   this   band of possible rotational velocities in   Weyl   gravity and test it against the family of the Milky Way rotation curves. To obtain the minimum rotation curve in Weyl gravity, we use the following values for the mass model parameters: $\Sigma^{0}_{thin} = 770.5$ $M_{\odot}pc^{-2}$, $R_{thin} =2.08 $ kpc, $\Sigma^{0}_{thick} = 97.8$ $M_{\odot} pc^{-2}$, $R_{thick}= 2.88$ kpc, $M_{bulge} = 1.7 \times 10^{10} M_{\odot}$ and $M_{bh} = 3.7 \times10^{6} M_{\odot}$. Similarly, we compute the maximum rotation curve using $\Sigma^{0}_{thin} = 1002.9$ $M_{\odot} pc^{-2}$, $R_{thin} = 3.12 $ kpc, $\Sigma^{0}_{thick} = 215.6$ $M_{\odot} pc^{-2}$, $R_{thick}= 4.32$ kpc, $M_{bulge} = 2.3 \times 10^{10} M_{\odot}$ and $M_{bh} = 4.3 \times10^{6} M_{\odot}$. The values of $M_{HI}$, $R_{HI}$, $M_{H2}$, $R_{H2}$ and $t$ remains same during this exercise as our mass model assumes only the mean values for these five parameters. We now choose five sets of updated rotation curves upto $\approx$ 100 kpc: BCK14 (a,b,c respectively) ([$R_0$,$V_0$] = [8.0 kpc, 200 km/s], [8.3 kpc, 244 km/s], [8.5 kpc, 220 km/s]; radially varying $\beta$) \cite{pijush}, YS12 ([$R_0$,$V_0$] = [8.0 kpc, 200 km/s]) \cite{sofue2} and YH15 \cite{yh16} ([$R_0$,$V_0$] = [8.34 kpc, 240 km/s]; radially varying $\beta$) \cite{yh16}. In Figure \ref{fig6}, we plot the rotational velocity band from the assumed mass model as well as rotation curve data linked to different sets of galactic constants. We see that the bracket easily encompasses the observed variations in the Milky Way rotation curve construction in all radial distances. Though the mean values of a couple of data points (15 kpc $< r <$ 25 kpc)  lie outside the bracket, their errorbars definitely fall within the allowed region and thus does not necessarily indicate a mismatch. This particular direction of analysis seals the success of Weyl gravity in explaining the observed Milky Way rotation curves. Furthermore, it shows that our results are robust against the existing uncertainty over galactic constants and anisotropy parameter.
\section{Investigating velocity dispersion of Globular Clusters}
\label{sec4}
\subsection{Velocity dispersion}
The velocity dispersion of galactic GCs is generally obtained as a function projected distance between the GC center and the stars (within GC) being observed. Such observations have unraveled an unusual feature in GC velocity dispersion. Velocity dispersion profiles have initially been observed to follow a monotonous decline similar to the Keplerian one. However, once it reaches a critical acceleration value $a_0$, generally associated with MOND regime, dispersion profile deviates from Newtonian expectation and shows a flattening trend. The asymptotic value of velocity dispersion varies from cluster to cluster.   This eventual flattening is difficult to comprehend in the context of Newtonian gravity (or GR) mainly because of the lack of dark matter in galactic globular clusters. One may therefore look for possible explanation in different modified gravity theories.  Here, we restrict ourselves to Weyl gravity only.\\

The Weyl gravity velocity dispersion profiles for galactic GCs could be easily derived using the Jeans equations. Almost spherical shape and isotropic dispersion profile of GCs indicate that they are spherically symmetric. Additionally, we assume GCs to be non-rotating. The
Jeans equation for the velocity dispersion $\sigma (r)$ thus takes the following form \cite{bt1987}:
\begin{equation}
	\frac{\partial(\rho\sigma^2)}{\partial r} +\frac{2 \rho(r) \beta \sigma^{2}(r)}{r} = - \rho(r)\frac{\partial\Phi}{\partial r},
	\label{eq:Jeans}
\end{equation}
where $r$ is the radial distance from the GC center, $\rho(r)$ is the radial density distribution function, and $\Phi(r)$ is the gravitational potential. We identify acceleration $a(r)=\partial\Phi/\partial r$  and utilize the constraint $\lim\limits_{r\rightarrow\infty}\sigma^2(r)=0$. Additionally, we assume anisotropy parameter $\beta=0$. Eq. (\ref{eq:Jeans}) thus gives
\begin{equation}
	\sigma^2(r)=\frac{1}{\rho}\int\limits_r^\infty\rho a(r')~dr'.
	\label{eq:sigma}
\end{equation}
\noindent Finally, the corresponding  projected line-of-sight (LOS) velocity dispersion reads   [see Eq. (14-16) in \cite{moffattoth}] :
\begin{equation}
	\sigma_\mathrm{LOS}^2(R)=\frac{\int_R^\infty r\sigma^2(r)\rho(r)/\sqrt{r^2-R^2}~dr}{\int_R^\infty r\rho(r)/\sqrt{r^2-R^2}~dr},
	\label{eq:LOS}
\end{equation}
where $R$ is the projected distance between the GC center and the stars being observed.
\subsection{Density distribution}
Because of the spherically symmetric nature, we can model GCs using simplistic Hernquist \cite{hern} profile
\begin{equation}
	\rho_\mathrm{hern}(r)=\frac{Mr_0}{2\pi r(r+r_0)^3},
\end{equation}
where $M$ is the total mass of the cluster, and $r_0$ is a characteristic radius. For GCs, we take the half-light radius as $r_0$. Our sample of GCs include NGC288, NGC1851, NGC1904 and NGC5139. Total luminosity and half-light radius for these globular clusters have been tabulated in Table \ref{T3}. Although, several other models can also be used to describe GCs, Moffat and Toth \cite{moffattoth} found that there is no significant impact of the choice of a particular mass model on the final velocity dispersion profile. For a spherically symmetric and extended mass distribution (like GCs), the acceleration in Weyl gravity can be obtained as \cite{weylcluster2}
\begin{align}
	&-\nabla\phi(r) = 
	\begin{aligned}[t]
		& G\Big[-{I_0(r)\over r^2} + {1\over R_0^2}\Big({I_2(r)\over 3 r^2} - {2\over 3} r E_{-1}(r)\\
		&- I_0(r)\Big) \Big] + {GM_0\over R_0^2}  - \kappa c^2 r,
	\end{aligned}
	\label{nc1}
\end{align}
where $I_n$ and $E_n$ are the interior and exterior moments of the mass profile defined respectively as  
\begin{equation}
	I_n(r)=4\pi \int_0^r \rho(x)x^{n+2} dx,
	\label{nc2}
\end{equation}
and
\begin{equation}
	E_n(r)=4\pi \int_r^{+\infty} \rho(x)x^{n+2}dx.
	\label{nc3}
\end{equation}
\noindent The constants $R_0$ ($= 24$~kpc) and $M_0$ ($= 5.6\times 10^{10}M_\odot$) replaces the usual  Weyl  gravity parameters $\gamma_{0}$ and $\gamma^{*}$. The first two terms in the Weyl gravity acceleration originates from the local mass distribution while the third term is the constant acceleration independent of the local source and linked to the universal Hubble flow. The fourth term, on the  other  hand, incorporates the effect of inhomogeneities on galactic or cluster motions.
\setlength{\tabcolsep}{12pt}
\begin{table}
	\caption{\label{T3}\textbf{Globular Cluster mass distribution profile}}
	\begin{ruledtabular}
		\begin{tabular}{ccc}
			& \textbf{half-light} & \textbf{Luminosity } \\
			& \textbf{radius (pc)} & \textbf{($L_{\odot}$)}\\
			& &\\
			\hline
			& &\\
			NGC 288 \cite{ngc288} & 2.9  & 3.9 $\times 10^{4}$ \\
			NGC 1851 \cite{ngc1851a1904} & 1.83  & 1.8 $\times 10^{5}$\\ 
			NGC 1904 \cite{ngc1851a1904} & 3.0 & 1.2 $\times 10^{5}$\\ 
			NGC 5139 \cite{ngcothers}& 7.7 & 1.1 $\times 10^{6}$ \\
		\end{tabular}
	\end{ruledtabular}
\end{table}
\subsection{Results}
\begin{figure*}[htb]
	\centering
	\subfigure[]{\label{fig:7a}
		\includegraphics[scale=0.33]{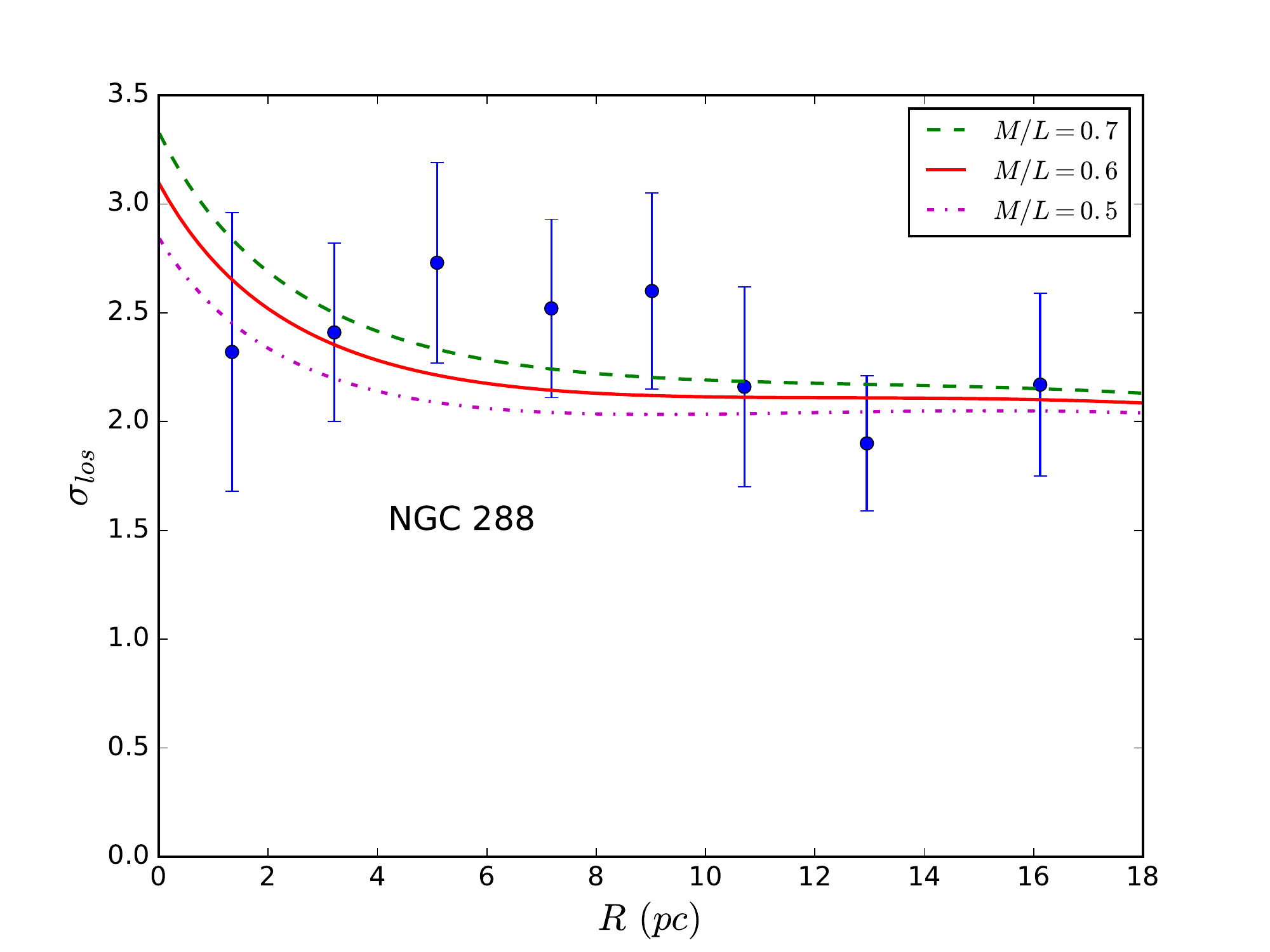}}
	\subfigure[]{\label{fig:7b}
		\includegraphics[scale=0.33]{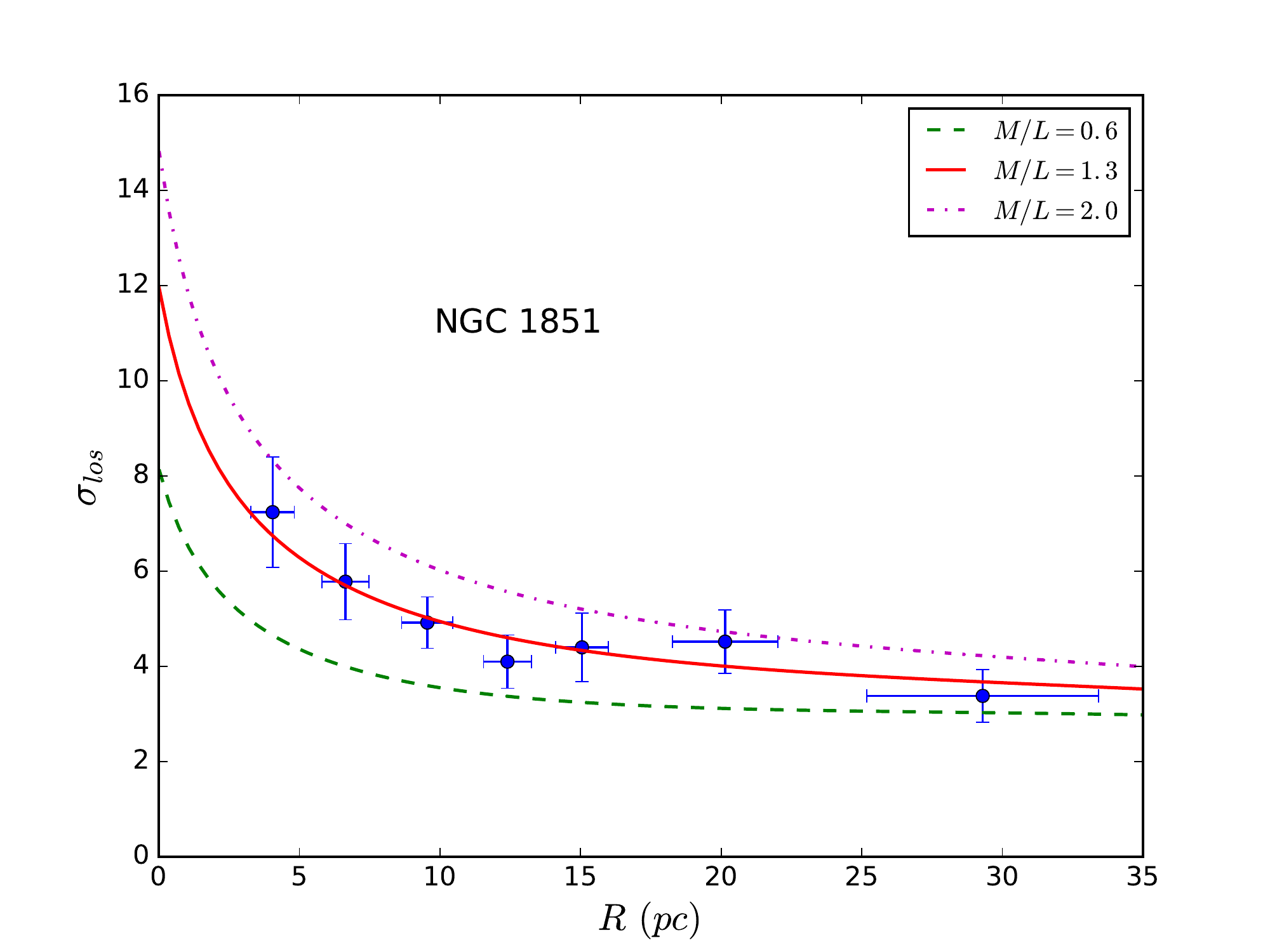}}
	\subfigure[]{\label{fig:7c}
		\includegraphics[scale=0.33]{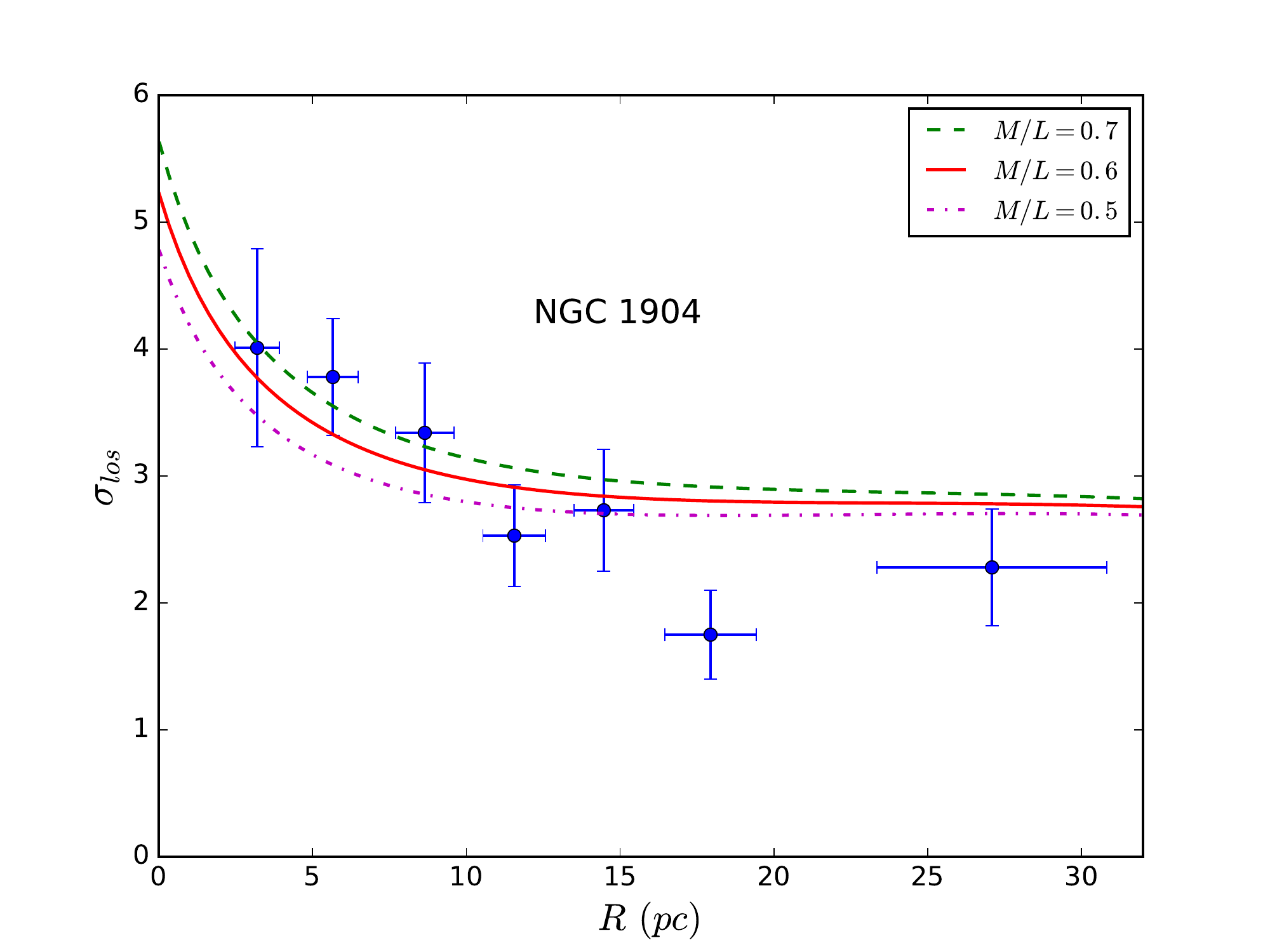}}
	\subfigure[]{\label{fig:7d}
		\includegraphics[scale=0.33]{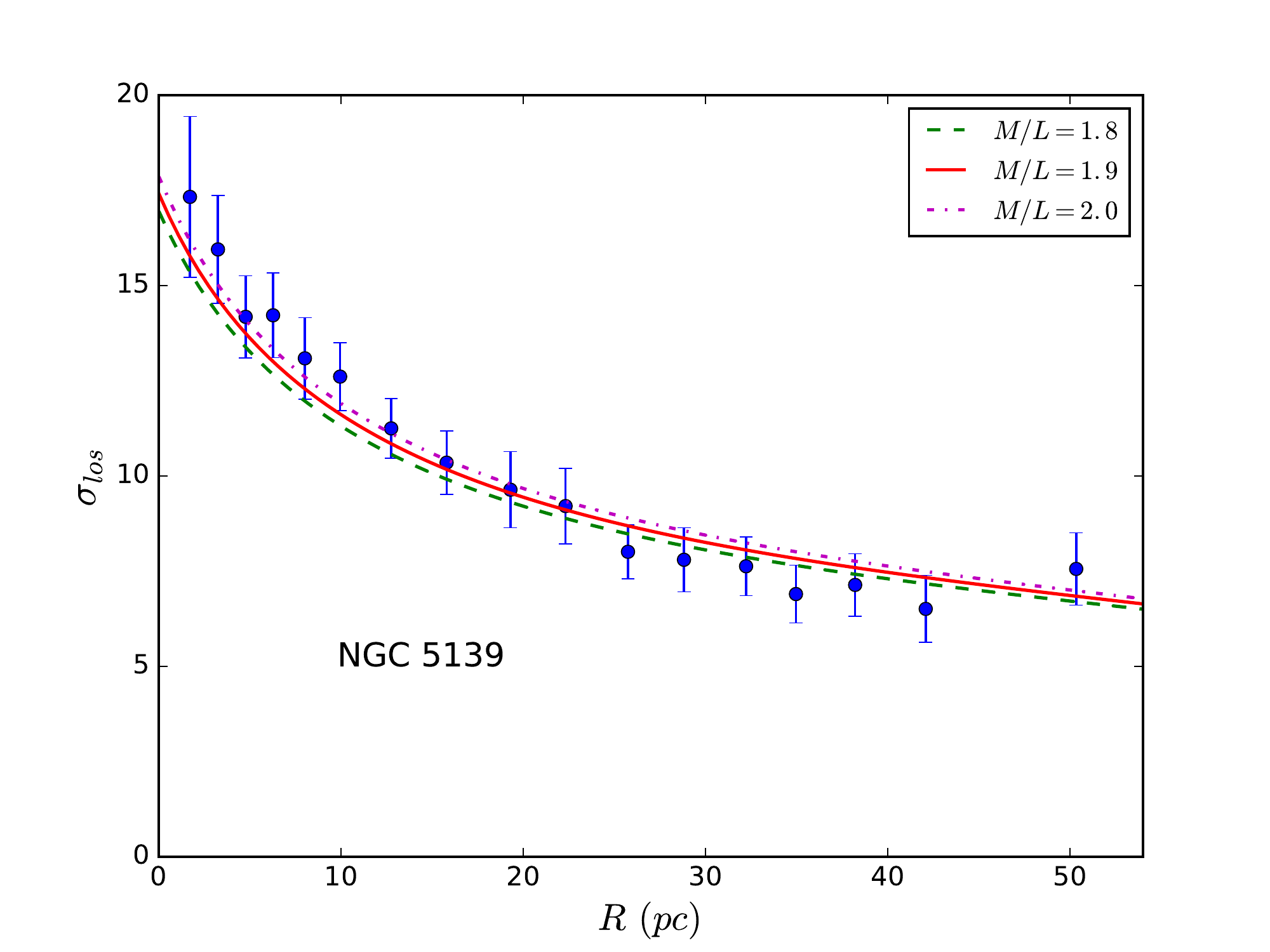}}
	\caption{\textbf{ \label{fig7} Observed elocity dispersions of four globular clusters have been plotted as a function of projected radial distances: (a) NGC288 (b) NGC1851 (c) NGC1904 (d) NGC5139. Weyl gravity predictions for different mass-to-light ratios have then been superimposed in green (dashed), red (lined) and violet (dash dotted). }} 
\end{figure*}
NGC 288 is a low concentration cluster and is located at a distance of 11.6 kpc from the galactic center. This particular globular cluster has an internal acceleration everywhere below the critical MOND value,  $a_0 = 1.14 \times 10^{-10} m/s^{2}$ . For this particular cluster, dispersion data is available up to 18 pc from the cluster center. Over this whole range of radii, the dispersion velocity is observed to be constant with an average value of 2.3 $\pm$ 0.15 km/s (Fig \ref{fig:7a}). For NGC 1851 (Fig \ref{fig:7b}), the velocity dispersion data extends up to 30 pc and converges towards a constant value 4.0 $\pm$ 0.5 km/s. The velocity dispersion fluctuates quite significantly due to the lack of sufficient amount of data. Still, the overall trend is prominent. In case of NGC 1904 (Fig \ref{fig:7c}), the dispersion velocities both increases and decreases a couple of times before settling down to a constant value of 2.25 $\pm$ 0.3 km/s. For this globular cluster, dispersion data covers the range of radii up to 30 pc from the cluster center. Beyond 10 pc, the dispersion fluctuates around the the mean constant value. Our final GC is NGC 5139 (Fig \ref{fig:7d}) or otherwise known $\omega$Cen. The cluster lies almost 6.4 kpc away from the Milky Way center and is one of the most massive clusters known. The dispersion profile of this cluster settles down to an asymptotic value of 7.0 km/s beyond a distance of 32 pc from the center of the cluster. Although the flattening is modest, it is easy to notice.\\ 

Rather than exactly fitting the velocity dispersion curve for these four GCs, our aim remains to qualitatively analyze the possibility of whether Weyl gravity can account for the observed dispersions. In our model, the only free parameter remains to be the mass-to-light ratio ($M/L$). We explore different values for $M/L$ to check whether an acceptable value can reproduce the dispersion data. Furthermore, this approach will also help us to understand the dependence of dispersion profile on the assumed $M/L$ values. For NGC 288, we find a good agreement with dispersion data with $M/L$=[0.5, 0.6, 0.7] (in solar unit). The resultant  Weyl  gravity dispersion profile becomes almost flat (with an asymptotic value of 2.2 km/s) beyond 6 pc from the center. The same set of values for $M/L$ yields similar degree of success for NGC 1904. However, the flattening in the predicted dispersion curve is subtle. In the case of NGC 1851, we obtain excellent fit with $M/L$=1.3. The predicted profile eventually settles to 3.9 km/s. We have also computed  Weyl  gravity predictions with $M/L$=0.6 and 2.0 and have found that they lie either below or above the observed values. However, the differences between predictions and observation is nominal. Furthermore, we successfully fit the dispersion profile of NGC 5139 with $M/L$ ratio 1.8, 1.9 and 2.0. In short, our analysis finds good fit with data with mass-to-light ratios in the range 0.5 $ < M/L < $ 2.0, which is consistent with previous photometric and population synthesis studies \cite{gcml1,gcml2,gcml3}.
\section{Confronting acceleration of Abell Cluster 1689}
\label{sec5}
\subsection{A1689 Baryonic Mass Profile}
\label{sec3a}
Galaxy clusters are the most massive gravitationally bound objects in the universe. In recent times, Abell 1689 (A1689) has caught enough interest among both modified gravity and dark matter proponents.  A1689 is located at a redshift z$=$0.1832. It is one of the largest and the most massive clusters ever observed.  The cluster has been extensively studied using weak and strong lensing, SZE and X-ray observations \cite{abel1,abel3}. These observations have estimated the galaxy and  gas contents  of A1689 with high accuracy.  Nieuwenhuizen  has recently inferred the acceleration profile for the cluster from the lensing data and claimed that modified gravity theories find it difficult to fit the A1689 acceleration data without assuming dark matter \cite{nieu1}. However, Moffat \& Haghighi \cite{moffat2017modified} have noted that the acceleration data could be well fitted by MOG while MOND is found not to comply with the inferred data. In a subsequent paper, Hodson \& Zhao \cite{mond3} have investigated the possibility to explain the inferred acceleration profile in two modified MOND frameworks without resorting to dark matter. In this section, we would look into the possibility to account for the acceleration data in the context of Weyl gravity.   \\
\setlength{\tabcolsep}{10pt}
\begin{table}[ht]
	\centering
	\caption[Baryon Mass Profile Parameters]{\textbf{Table of parameters for the galaxy mass profile and the gas profile as taken from \cite{nieu1} and \cite{abel1}.}}
	\label{T4}
	\begin{tabular}{c  c  c}
		&  &  \\ 
		Parameter & Value & Unit \\
		&  &  \\
		\hline \\
		$M_{cg}$ & $3.2\times 10^{13}$ & $M_{\odot} $             \\
		$R_{co}$ & $5$                         & kpc                       \\
		$R_{cg}$ & $150$                       & kpc                       \\
		$n_{e0}$ & $0.0673$                    & cm$^{-3}$ \\
		$R_{g}$  & $21.2$                      & kpc                       \\
		$n_{g}$  & $2.91$                      & n/a                       \\
		$k_{g}$  & $1.9$                       & n/a  \\
		&  & \\
		\hline                   
	\end{tabular}
\end{table}

We now present the baryonic mass model for the cluster, A1689. We assume the cluster to be spherically symmetric \cite{abel3}. Generally, clusters consist of two main sources of baryonic mass: galaxies and intra-cluster gas. While cluster center is dominated by galaxies, gas dominates the outer region. Though there are many galaxies in the cluster, we assume that the galaxy mass density of A1689 is dominated by the Brightest Cluster Galaxy (BCG) residing in the center of the cluster and will extrapolate the corresponding mass density over the whole cluster. We use the BCG density profile proposed by  Limousin et. al. \cite{abel1}:
\begin{equation}
	\rho_{gal}(r)=\frac{M_{cg} (R_{co} + R_{cg})}{2 \pi^2 (r^2+R^2_{co})(r^2+R^2_{cg})} , 
\end{equation}
where  $M_{cg}$ and $R_{cg}$  are the mass and radial extent of the central galaxy, respectively, while  $R_{co}$  is the core size. This BCG profile does not include any dark matter contribution and is solely linked to the stellar contents of the galaxies. For the gas,  we use a cored Sersic electron density profile obtained from CHANDRA X-ray observations \cite{nieu2}:
\begin{equation}
	n_e(r)=n^0_e exp \left[ k_g - k_g \left(1 + \frac{r^2}{R^2_{cg}} \right)^\frac{1}{2 n_g} \right] , 
\end{equation}
and
\begin{equation}
	\rho_{gas}(r)=1.167 m_N n_e(r) , 
\end{equation}
where $n^0_e$ is the central electron number density and $R_g$ is the radial extent of the intra-cluster gas. $k_g$ and $n_g$ controls the shape of the density profile. Both the galaxy and gas mass density profiles considered in our work have previously been employed in \cite{moffat2017modified} and \cite{mond3}. Values for different mass profile parameters used in this work have been listed in Table \ref{T4}.\\
\begin{figure}[h]
	\begin{center}
		\includegraphics[scale=0.4]{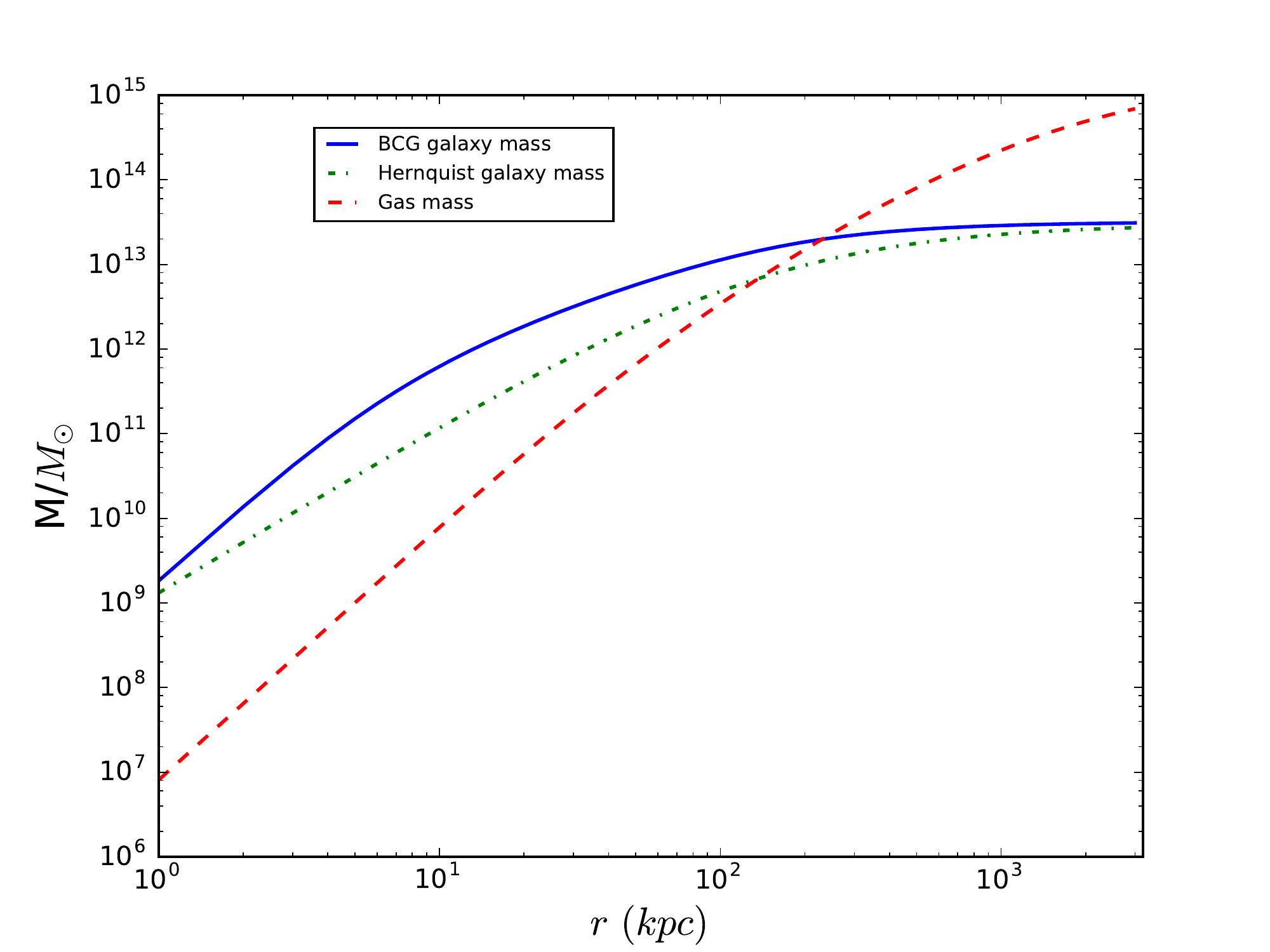} 
	\end{center} 
	\caption{\textbf{Plot shows the total galaxy mass profile of cluster A1689 for both the mass model: BCG model (in blue) \& Hernquist model (in green). The total gas mass of the cluster has then been plotted in red  (dashed). It is easy to notice that galaxy mass dominates in the central part and gas becomes dominant in the outer region.}}
\end{figure}

The   total baryonic mass density of the cluster is now 
\begin{equation}
	\rho_{bar}(r)=\rho_{gal}(r)+\rho_{gas}(r)
\end{equation}
and total baryonic mass $M_{bar}$ could be obtained by integrating $\rho_{bar}$ over the volume of the cluster
\begin{equation}
	M_{bar}(r)=4 \pi \int^{r}_{0} \rho_{bar}(r') r'^{2}dr.\\
\end{equation}

We note that Hodson \& Zhao \cite{mond3} also used  an empirical Hernquist galaxy mass profile for A1689 . They argued that this particular model exhibits a desired behaviour for the baryons in the interior of the cluster within modified gravity regime. Though this particular mass profile is obtained empirically, we use it to see whether choosing a different galaxy mass profile will alter Weyl gravity acceleration significantly. The Hernquist profile \cite{hern} is described by\begin{equation}
	M_{H}(r) = \frac{M_{h} r^{2}}{\left( r + h \right)^{2}},
\end{equation}
where $M_{h}$ and $h$ are the total mass and radial extent of the galaxies respectively. $M_{h}$ has been taken as $3\times 10^{13}$ $M_{\odot}$ while $h$ has been set to 150 kpc (following \cite{mond3}). \\

Weyl gravity acceleration could now be readily obtained from the total baryonic mass profile as [ similar expression used earlier for GCs; Eq. (\ref{nc1}) ]:
\begin{align}
&a_{cluster} = 
\begin{aligned}[t]
& G\Big[-{I_0(r)\over r^2} + {1\over R_0^2}\Big({I_2(r)\over 3 r^2} - {2\over 3} r E_{-1}(r)\\
&- I_0(r)\Big) \Big] + {GM_0\over R_0^2}  - \kappa c^2 r,
\end{aligned}
\label{eq10}
\end{align}
where $I_n(r)$ and $E_n(r)$ are interior and exterior mass moments defined in Eq. (\ref{nc2}) and Eq. (\ref{nc3}) respectively.  
\subsection{A1689 acceleration data}
To investigate  the acceleration profile of A1689  in the light of Weyl gravity, we take the following approach. First, we infer the acceleration data from strong and weak lensing observation. For that, we adopt the method prescribed by Nieuwenhuizen \cite{nieu1}. This approach requires the knowledge of two observables:  surface mass density $\Sigma (r)$ (strong lensing) and transversal shear $g_t(r)$  (weak lensing). For strong lensing, surface mass density is related to acceleration through
\begin{equation}
	a(r)\lesssim 2\pi G\Sigma(r),
\end{equation}
where $G$ is the Newtonian Gravitational constant.  In case of weak gravitational lensing, one can obtain a similar relation between acceleration and transversal shear:
\begin{equation}
	a(r)\lesssim 2\pi G\Sigma_c g_t(r),
\end{equation}
where $\Sigma_c$ is the critical surface mass density. It must be noted that this acceleration is approximate in nature  and indicates  the upper limit of the acceleration. However, it would still provide a good estimate for acceleration profile of A1689 and thus could be used as a probe for modified gravity theories like Weyl gravity. Simultaneously, we would compute the acceleration within the context of Weyl gravity from the baryonic mass model  presented in Section \ref{sec3a} using Eq. (\ref{eq10}). We would like to see whether the computed acceleration from baryons matches with the inferred profile from lensing data or not. 
\begin{figure}[h]
	\centering
	\includegraphics[width=1.0\linewidth]{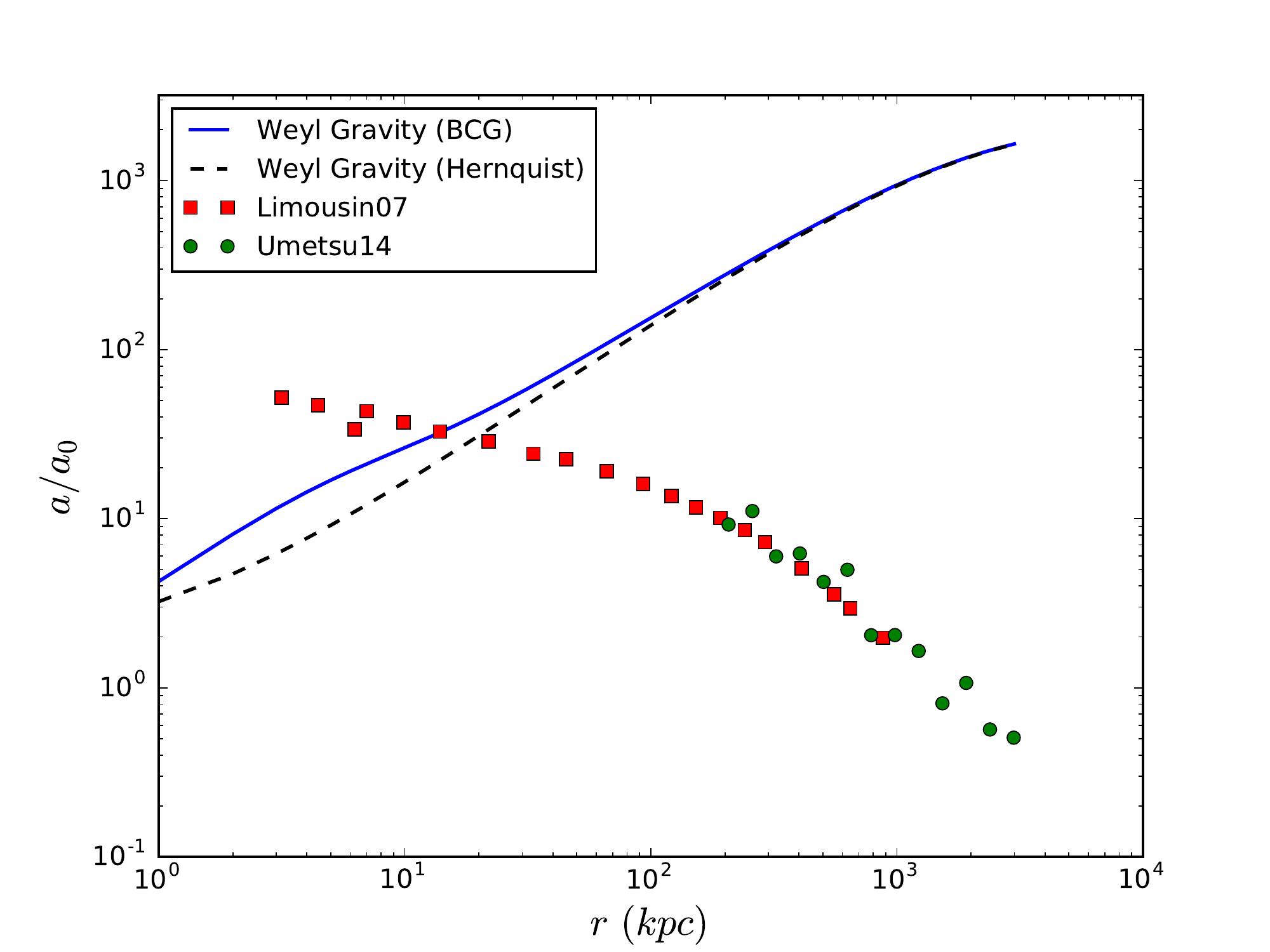}
	\caption[Weyl Gravity Acceleration for $R_0$=24 kpc]{\textbf{Plot shows the predicted Weyl gravity acceleration in A1689 for both BCG galaxy model  (blue, lined) and Hernquist model (black, dashed)  (in loglog scale). Acceleration data derived from strong lensing analysis of Limousin et al (Limousin07) and weak lensing analysis of Umetsu et al (Umetsu14) is plotted in red square and green circles respectively.}}
	\label{fig:9}
\end{figure}
\subsection{Results}
We first obtain the acceleration profile combining strong lensing (SL) observation from Limousin et al \cite{abel1} and weak lensing (WL) data from Umetsu et al \cite{abel3}. While  the  data from \cite{abel1} extends upto 500 kpc from the center of the cluster, WL data \cite{abel3} covers the range of radii in between 200 kpc and 3000 kpc. Thus a combination of SL and WL data helps us to probe both the interior and exterior of the cluster. We now present the normalized inferred acceleration in Figure \ref{fig:9}. The  inferred  acceleration  is found to be steadily decreasing.  The transition between  strong acceleration ( $>$ $10^{-10} m/s^{2}$) and weak acceleration  ( $<$ $10^{-10} m/s^{2}$)  regime is smooth and featureless.\\

Up next, we use the baryonic mass model of galaxies and gas of the cluster to compute the Weyl gravity acceleration using Eq. (\ref{eq10}). We use the parameter values $R_0$=24 kpc and $M_0$=5.6 $\times 10^{10} M_{\odot}$ obtained earlier from the Galaxy rotation curve studies \cite{weylrot1,weylrot2,weylrot3,weylrot4}. The resultant acceleration profile shows an increasing trend contrary to the fall-off exhibited in the profile inferred from lensing data (Figure \ref{fig:9}). The discrepancy between inferred acceleration (from lensing) and computed Weyl gravity acceleration (from baryons) increases with increasing radial distances from the center of the cluster. At the outskirts (r$\sim$1 Mpc), Weyl gravity predictions are found to be, on an average, almost two to three orders of magnitude higher than the inferred acceleration.\\ 

We note that similar kind of analysis with Weyl gravity has also been done by Horne \cite{weylcluster2} for A2029. Using CHANDRA X-ray data, Horne showed that the total mass (within the radial distance of 300 kpc) obtained through integrating the X-ray gas density profile is roughly 10 times more than the mass one would expect in Weyl gravity given hydrostatic equilibrium is maintained. The acceleration profile obtained from X-ray hot gas hydrostatic equilibrium for A2029 \cite{weylcluster2} (Thick black curve in Fig. 1(f) of \cite{weylcluster2}) and acceleration obtained through lensing for A1689 \cite{nieu1}, within the radial distance of 300 kpc,  (red squares and green dots in Fig. \ref{fig:9}) look strikingly similar. Both have a slowly decreasing trend beyond 30 kpc from cluster center.  In both the cases, derived acceleration from total reported baryon mass (compare black thin continous line in Fig. 1(f) of \cite{weylcluster2} with blue/black continous/dashed line in Fig. \ref{fig:9}) is found to increase, and becomes stronger than either  the value extracted from hydrostatic equilibrium (Fig. 1(f) of \cite{weylcluster2}) or the  acceleration inferred from lensing (Fig. \ref{fig:9}). Our result is therefore similar to what Horne found out: conformal gravity becomes stronger in the cluster scale. \\

We also find that the choice of galaxy profiles does not alter our conclusion. However, acceleration predicted in BCG model is higher in the central part of the cluster. This difference becomes negligible in the outer part of A1689 (beyond 300 kpc from the cluster center). Furthermore, we have considered the possibility where Weyl gravity parameters $R_0$, $M_0$ and $\kappa$  might be allowed to take different values at the scale of clusters. We note that the overall effect of $M_0$ in the acceleration profile will not be significant. A higher value of $\kappa$ could have been effective to arrest the increasing acceleration (from baryons) at the outer part of the cluster. However, there is no apparent reason for these parameters to assume different values at extra-galactic scales. \\

One could also derive the required enclosed mass for A1689 in Weyl gravity from the radial profile of acceleration obtained from lensing (and from hydrostatic equilibrium) alone. The non-local nature of gravity in conformal theory will make this work difficult though. Additionally, the immediate non-availability of the data for the radial profiles of temperature, density and pressure in the gas is an issue. Though we agree that such an analysis could expand the scope of this paper, it is beyond the illustrative purpose of this work. We thus leave this piece of analysis for future explorations. However, as the acceleration generated from the reported baryon mass in the cluster is found to exceed the inferred acceleration, one can expect that the enclosed mass required to fit the acceleration profile would be smaller than the reported baryon mass. Thus, the discrepancy would then shape itself in terms of total enclosed mass (as reported in \cite{weylcluster2}).\\

At this point, it is crucial to note that, in Weyl gravity, the local dynamics is influenced by both the local and global distribution of matter. To account for the global contribution to the gravity within the cluster, we have included two global terms having different origins. However, the effect of the nearby external matter may not be well captured by these global terms. In fact, the contribution from nearby external mass distribution (particularly from low density voids) could in principle result a shielding mechanism and may potentially reduce the gravity of the interior matter \cite{weylcluster2}. Incorporating such effects might not be straight-forward given the highly non-local nature of the gravitational field and is beyond the scope of this paper.\\ 

\section{Discussion \& Conclusions}
\label{sec6}
We have tested Weyl gravity from galactic scale up to the scale of galaxy clusters. At galactic scale, we test the viability of Weyl gravity with the extended rotation curve of the Milky Way and velocity dispersions of four globular clusters. In our quest to find clues for modified gravity in the Milky Way rotation curve, we first identified that, as predicted in Weyl gravity, the rotation curve indeed falls at larger distances. We have demonstrated that including a central supermassive blackhole in the mass model improves the Weyl gravity predictions manifold. Furthermore, we find Weyl gravity predictions to be consistent with radial acceleration relation (RAR) in the Milky Way. Additionally, we compute a bracket of rotational velocities possible in the Milky Way within the context Weyl gravity and found that this bracket accommodates the diverse  rotation curves for the Milky Way, which is a result of inherent assumptions for the different values of $R_0$,  $V_0$  and $\beta$, made during the construction of rotation curve profile. In our analysis, we have used rotation curve data from three different groups:  BCK14, YS12 and YH16 \cite{pijush,sofue2,yh16}. The range of values for several galactic constants assumed in these studies are: $8.0$ kpc $<$ $R_0 < 8.5$ kpc; 200 km/s $< V_0 <$ 244 km/s and 0 $< \beta <$ 1. Thus, rotational curve data used in our work truly represents the family of the MW rotation curves. Our result is therefore immune to current observational errors and uncertainties. This study thus is not only different from previous Weyl gravity analysis of the Milky Way rotation curve \cite{obrien}, it is actually complementary to that.\\

In  the  case of GCs, we have calculated the velocity dispersions for NGC 288, NGC 1851, NGC 1904 and NGC 5139. We assumed the GCs to be spherically symmetric and non-rotating. Furthermore, we adopted simple Hernquist mass profile for GCs. However, in reality, NGC 1851, NGC 1904 and NGC 5139 has been observed to be slowly rotating. Moreover, our analysis does not include any complicated tidal effects or external field effects due to the gravitational pool of the Milky Way. Still, we find good fits to the observed dispersion profiles with reasonable values of mass-to-light ratio. We note that Moffat \& Toth have obtained a similar fit within the context of MOG with $M/L=4.38$ for NGC 288 and $M/L=2.79$ for NGC 5139 \cite{moffattoth}. On contrary, our analysis results 0.5 $< M/L <$ 2.0. This range of mass-to-light ratio is more consistent with recent estimates \cite{gcml1,gcml2,gcml3}. On top of that, our sample of GCs are extremely diverse. They have different sizes, different luminosities, different concentrations, different dynamical histories and they lie at different radial distances from the galactic center. Thus, they experience different strength of gravitational pull. Still, simple Weyl gravity model can more-or-less describe their dispersion profile which is otherwise difficult to explain in Newtonian dynamics (or in GR). Such universal explanation for the eventual flattening of dispersion profiles in GCs should definitely be taken as a triumph for modified gravity and in particular for Weyl gravity.\\

We have then extended our study to Abell cluster 1689 (A1689). For A1689, we modelled the galaxy cluster in Weyl gravity and compared the results with inferred acceleration profile from lensing data. Weyl gravity acceleration has been found to keep increasing with distances from the center of the cluster and exceed the inferred profile by almost two to three orders of magnitude in the outer region (beyond 300 kpc). The essence of our result is similar to the claims of Horne \cite{weylcluster2} and Diaferio \& Ostorero \cite{weylcluster1}. Horne \cite{weylcluster2} found that Weyl gravity analysis of X-ray gas in Abell 2029 yields a total mass profile which is nearly 10 times greater than what is required to hold the hot gas in hydrostatic equilibrium. Such disagreement with observation has then further extended to temperature profile by Diaferio \& Ostorero who used adiabatic N-body/hydrodynamical simulations of isolated self-gravitating gas clouds in galaxy clusters within the framework of Weyl gravity and noted that the predicted temperature profile rises, rather than following a decreasing trend observed in real clusters. It suggests that the success of Weyl gravity at the galactic scale does not get translated in the scale of clusters.\\

However, we note that, in dark matter formalism, the acceleration (or velocity) is determined almost by dark matter distribution. Thus, a little uncertainty in baryonic mass does not affect the overall expectation. That is not the case for modified gravity theories like Weyl gravity. As the observed acceleration (or equivalently velocity) is completely determined by the visible baryonic mass distribution, extra caution must be taken while adopting a particular mass model. It is worth pointing out that the presence of foreground and background structures in the line of sight of A1689 can increase the uncertainty in the estimated mass (from lensing data) \cite{1689a}. Even any departure from spherical symmetry will have a similar effect \cite{1689b}. However, even if these factors have somehow contributed to the uncertainty of the mass profile used, it is highly unlikely that they will severely alter our result. Furthermore, the inferred acceleration data is no way an explicit acceleration profile. It shows a trend similar to the ones observed in several galaxies. Therefore, the inferred profile may be a good estimate for the actual centripetal acceleration profile. Still, it is not clear whether that is indeed the case. Existence of several structures aligned along the line of sight makes kinematic studies difficult at present \cite{1689c}. On a more theoretical ground, the appropriate inclusion of the shielding effects of nearby external matter of the cluster could help Weyl gravity to reconcile with inferred acceleration profile. However, such effects are currently poorly understood in Weyl gravity. Thus much more work is required in both Weyl gravity as well as kinematic studies of A1689 before reaching any strong conclusion and is left for future.\\

Before we conclude, we would like to point out an generally overlooked but important aspect of Weyl gravity.  Weyl gravity, like all other fourth order gravity theories, does not possess any dimensional constant. Instead, it features a dimensionless constant $\alpha_{g}$ which has a value of order unity. However, when Weyl gravity is coupled with matter, the presence of a dimensional constant (namely Newtonian gravitational constant G) is assumed. There lies some well supported motivations behind such exercise. In fact, such dimensional constant is shown to be induced by different interactions in (quantum) Weyl gravity \cite{zee1983einstein}.\\

In summary, we have demonstrated that Weyl gravity can achieve high degree of success in describing the observed rotation curves of the Milky Way without invoking any dark matter profile. Our study has then extended the credibility of Weyl gravity to the scale of globular clusters. However,  the Weyl gravity acceleration generated from the reported baryon mass in the cluster is found to exceed the inferred acceleration from lensing data. This apparent discrepancy  may in principle be tackled by properly including the effects of the nearby external matter. This particular avenue of research  needs  to be explored further before reaching a final conclusion.
\begin{acknowledgements}
We would like to thank Pijushpani Bhattacharjee, Sourya Ray and Rajesh Nayak for fruitful discussions. We are also grateful to Keiichi Umetsu, Marceau Limousin and Pijushpani Bhattacharjee for providing us the necessary data. K.D. is partially supported by a Ramanujan Fellowship under SERB, DST, Govt. of India.  K.D.  would like to thank the Abdus Salam International Centre for Theoretical Physics, Trieste for hospitality when the final stages of the work were completed.
\end{acknowledgements}
\bibliographystyle{apsrev4-1}
%
\end{document}